\documentclass{ieeeaccess}
\usepackage{cite}
\usepackage{amsmath,amssymb,amsfonts}
\usepackage{array}
\usepackage{algorithmic}
\usepackage{graphicx}
\usepackage{textcomp}
\usepackage{matlab-prettifier}
\usepackage{listings}
\def\BibTeX{{\rm B\kern-.05em{\sc i\kern-.025em b}\kern-.08em
		T\kern-.1667em\lower.7ex\hbox{E}\kern-.125emX}}

\usepackage{mathtools}
\DeclarePairedDelimiter\ceil{\lceil}{\rceil}
\DeclarePairedDelimiter\floor{\lfloor}{\rfloor}
\DeclarePairedDelimiter\round{\lfloor}{\rceil}
\DeclarePairedDelimiter\abs{\lvert}{\rvert}%
\DeclareMathOperator{\sech}{sech}
\DeclareMathOperator*{\maxi}{max}
\DeclareMathOperator*{\mini}{min}
\DeclareMathOperator*{\expm}{expm}
\DeclareMathOperator*{\FLOPs}{FLOPs}
\usepackage{array}
\usepackage{multirow}
\newcolumntype{P}[1]{>{\centering\arraybackslash}p{#1}}
\newcolumntype{M}[1]{>{\centering\arraybackslash}m{#1}}
\makeatletter
\renewcommand*\env@matrix[1][*\c@MaxMatrixCols c]{%
	\hskip -\arraycolsep
	\let\@ifnextchar\new@ifnextchar
	\array{#1}
}
\makeatother

\begin{document}
\history{Date of publication xxxx 00, 0000, date of current version xxxx 00, 0000.}
\doi{10.1109/ACCESS.2017.DOI}

\title{Fast Nonlinear Fourier Transform Algorithms Using Higher Order Exponential Integrators}
\author{\uppercase{Shrinivas Chimmalgi}\authorrefmark{1},
		\uppercase{Peter J. Prins}\authorrefmark{1}, and \uppercase{Sander Wahls}\authorrefmark{1}
		\IEEEmembership{Senior Member, IEEE}}
\address[1]{Delft Center for Systems and Control, Delft University of Technology, Mekelweg 2, 2628 CD Delft, The Netherlands.}
\tfootnote{This project has received funding from the European Research Council (ERC) under the European Union's Horizon 2020 research and innovation programme (grant agreement No 716669).}

\markboth
{Author \headeretal: Preparation of Papers for IEEE TRANSACTIONS and JOURNALS}
{Author \headeretal: Preparation of Papers for IEEE TRANSACTIONS and JOURNALS}

\corresp{Corresponding author: Shrinivas Chimmalgi (e-mail:  s.chimmalgi@tudelft.nl).}

\begin{abstract}
The nonlinear Fourier transform (NFT) has recently gained significant attention in fiber optic communications and other engineering fields. Although several numerical algorithms for computing the NFT have been published, the design of highly accurate low-complexity algorithms remains a challenge. In this paper, we present new fast forward NFT algorithms that achieve accuracies that are orders of magnitudes better than current methods, at comparable run times and even for moderate sampling intervals. The new algorithms are compared to existing solutions in multiple, extensive numerical examples.
\end{abstract}


\titlepgskip=-15pt

\maketitle

\section{Introduction}

\IEEEPARstart{T}{he} fast Fourier transform (FFT) is a well-known success story in engineering. From a numerical point of view, the FFT provides a mere first-order approximation of the discrete-time Fourier transform one is actually interested in. Hence the success of the FFT is quite surprising. Upon closer inspection, it however turns out that approximations based on FFTs are very accurate if the signal is smooth \cite{Epstein05}. Recently, \emph{nonlinear} Fourier transforms (NFTs) have been gaining much attention in engineering areas such as fiber-optic communications \cite{Review1, Review2} and coastal engineering \cite{osborne2018nonlinear,Bruhl2014}. NFTs are generalizations of the conventional Fourier transform that allow to solve specific nonlinear evolution equations in a way that is analogous to how Fourier solved the heat equation \cite{AKNS}. The evolution of the nonlinear Fourier spectrum is, exactly like in the linear case, much simpler than the evolution of the original signal. NFTs also have unique data analysis capabilities that enable the detection of particle-like signal components known as solitons \cite{Singer1999}. 

Recently, a nonlinear variant of the FFT has been derived \cite{PF,FNFT}. These type of fast NFTs (FNFTs) can provide up to second-order accuracy \cite{Review2}. Unfortunately, unlike for the FFT, the accuracy of the FNFTs in \cite{PF,FNFT,Review2} remains (at most) second-order even when the signal is smooth. As a result, engineers currently have to strongly oversample even smooth signals in order to get reliable numerical results \cite[Section 4]{Civelli2018}. Several authors have proposed NFT algorithms with higher orders of accuracy, utilizing either Runge-Kutta \cite{RK4,IT_NFT2} or implicit Adams methods \cite{IA2}. However, even though these methods have higher accuracy orders, they require very small sampling intervals in order to actually perform better than standard second-order method such as \cite{BO}. For practically relevant sampling intervals, they are typically not the best choice as they are slower \emph{and} may even perform worse in these regimes. Numerical methods that provide better complexity-accuracy trade-offs in practically relevant regimes have been an open problem until recently. 

In \cite{CF4_APC}, the authors introduced a new numerical method that can compute the NFT with accuracies that are orders of magnitudes better than those of standard methods while having comparable run times. The key enabler for this large improvement in the complexity-accuracy trade-off was that, for the first time, a so-called commutator-free exponential integrator \cite{CF4_2} of higher order was used to compute the NFT. In a nutshell, the absence of commutator terms drastically reduces the computational cost whereas the excellent performance of exponential integrators is retained. However there is one drawback remaining in \cite{CF4_APC}: The complexity of the algorithm grows quadratically with the number of signal samples $D$, which makes the algorithm attractive only if the number of samples is not too high. In other words the algorithm is not fast. In this paper we overcome this limitation. \textbf{Our main contribution is the first fast higher-order NFT algorithm based on an exponential integrator. By combining it with Richardson extrapolation scheme, we arrive at an NFT algorithm that requires only $\mathcal{O}(D\log^2D)$ floating point operations, but achieves a sixth-order [$\mathcal{O}(D^{-6})$] error decay.\footnote{The complexity estimate only contains the cost of computing the so-called continuous spectrum as is usual in the NFT literature. Details on the continuous spectrum will be given later in the text. The cost of computing the discrete spectrum are highly problem dependent.} To the best of our knowledge no such algorithm has been investigated in the literature before. We show that the complexity-accuracy trade-off of the proposed algorithm is dramatically better than that of existing standard methods.} To give an illustration, we point out that in one of our numerical examples, our new method achieves an accuracy that is \emph{hundred million} times better than the standard second-order method in \cite{BO} at a comparable run time.\footnote{Compare the error for CF$_1^{[2]}$ in Fig. \ref{fig:CF_sech_focusing_tradeoff} with that of FCF\_RE$_2^{[4]}$ in Fig. \ref{fig:FCFvsFCF_RE_sech_focusing_tradeoff} for the execution time $1$ sec. We remark that although the execution times are implementation specific, they still give a good indication of the advantages of our proposed algorithm (see Appendix A).} 

The rest of this paper is structured as follows. In Section \ref{Preliminaries} we recapitulate the required mathematical background of the NFT. In Section \ref{NFT} we derive improved versions of our recently proposed numerical NFT in \cite{CF4_APC}, and compare them with both conventional second-order and other higher-order NFT algorithms in multiple numerical examples. Then, in Section \ref{FNFT}, we demonstrate how some of our new numerical NFTs can be made fast. The fast versions are compared to their slow counterparts. Next, in Section \ref{Series_acceleration}, we investigate how Richardson extrapolation can improve the complexity-accuracy trade-off of the fast NFT methods even further. The paper is finally concluded in Section \ref{conclusion}.\footnote{Some of the results were presented at the OSA Advanced Photonics Congress, Zurich, July 2018 (SpM4G.5)}%

\subsection*{Notation}
Real numbers: $\mathbb{R}$; $\mathbb{R}_{\geq 0}\ :=\ \{x\in\mathbb{R}:x\geq 0\}$; Complex numbers: $\mathbb{C}$; Complex numbers with positive imaginary part: $\mathbb{H}$; Integers: $\mathbb{Z}$; $i:=\sqrt{-1}$; Euler's number: e; Real part: $\text{Re}(\cdot)$; Complex conjugate: $(\cdot)^*$; Floor function: $\floor{\cdot}$; Absolute value: $\abs{\cdot}$; Matrix exponential: expm$(\cdot)$; Matrix product: $\prod_{k=1}^{K}A_k:=A_KA_{K-1}\times\cdots\times A_1$; Matrix element in the $i$th column and $j$th row: $[\cdot]_{i,j}$; Fourier transform of the function $f(t)$, $\mathcal{F}(f(t))= \tilde{f}(\xi)=\int_{-\infty}^{\infty}f(t)e^{-i t \xi}dt$; Inverse Fourier transform of the function $\tilde{f}(\xi)$, $\mathcal{F}^{-1}(\tilde{f}(\xi))=f(t)=\frac{1}{2\pi}\int_{-\infty}^{\infty}\tilde{f}(\xi)e^{i t \xi}d\xi$.

\section{Preliminaries}\label{Preliminaries}
In this section we describe the mathematical machinery behind the nonlinear Fourier transform (NFT). For illustration purposes we will describe the NFT in the context of fiber-optic communications. Let $q(x,t)$ denote the complex envelope of the electric field in an ideal optical fiber, whose evolution in normalized coordinates is described by the nonlinear Schr\"odinger equation (NSE) \cite[Chap. 2]{GPA}
\begin{equation}
i\frac{\partial q}{\partial x}+\frac{\partial^2 q}{\partial t^2}+2\kappa \lvert q \rvert^2q=0.
\label{eqn:NSE}
\end{equation}
 Here, $x\geq0$ denotes the location in the fiber and $t$ denotes retarded time. The parameter $\kappa$ determines if the dispersion in the fiber is normal (-1) or anomalous (+1). When $\kappa=+1$, \eqref{eqn:NSE} is referred to as the focusing NSE and for $\kappa=-1$ \eqref{eqn:NSE} is referred to as the defocusing NSE. The NFT that solves the NSE \eqref{eqn:NSE} is due to Zakharov and Shabat \cite{ZS}. It transforms any signal $q(t)$ that vanishes sufficiently fast for $t\to\pm\infty$ from the time-domain to the nonlinear Fourier domain through the analysis of the linear ordinary differential equation (ODE)
\begin{equation}
\frac{\partial V(t,\lambda)}{\partial t}=C(t,\lambda) V(t,\lambda)=\begin{bmatrix}
-i\lambda &q(t)\\-\kappa q^*(t) &i\lambda
\end{bmatrix} V(t,\lambda),
\label{eqn:ZS}
\end{equation}
where $q(t)=q(x_0,t)$ for any fixed $x_0$, subject to the boundary conditions
\begin{equation}
\begin{aligned}
V(t,\lambda)=&
\left[\phi(t,\lambda) \ \bar{\phi}(t,\lambda)\right]
\to\begin{bmatrix}[c<{\mspace{-5mu}}>{\mspace{-5mu}}c]
e^{-i\lambda t}&0\\0&-e^{i\lambda t}
\end{bmatrix} \text{as} \ t\to -\infty,\\
V(t,\lambda)=&
\left[\bar{\psi}(t,\lambda) \ \psi(t,\lambda)\right]
\to\begin{bmatrix}[c<{\mspace{1mu}}>{\mspace{1mu}}c]
e^{-i\lambda t}&0\\0&e^{i\lambda t}
\end{bmatrix} \text{as} \ t\to \infty.
\end{aligned}
\label{eqn:Jost}
\end{equation}
The term $\lambda\in \mathbb{C}$ is a spectral parameter similar to $s$ in the Laplace domain. The matrix $V(t,\lambda)$ is said to contain the eigenfunctions since \eqref{eqn:ZS} can be rearranged into an eigenvalue equation with respect to $\lambda$ \cite{AKNS}. One can view the eigenfunctions $V(t,\lambda)$ as being scattered by $q(t)$ as they move from $t\to -\infty $ to $t\to \infty$. Hence \eqref{eqn:ZS} is referred to as the scattering problem \cite{ZS}. (Many problems in signal processing can be expressed through such scattering problems \cite{inverse_scattering_framework}.) For \eqref{eqn:ZS} subject to boundary conditions \eqref{eqn:Jost}, there exists a unique matrix
\begin{equation}
\mathcal{S}(\lambda) =\begin{bmatrix}
a(\lambda) &\bar{b}(\lambda)\\
b(\lambda) &-\bar{a}(\lambda)
\end{bmatrix},
\label{eqn:Scattering_matrix}
\end{equation}
 called the scattering matrix, such that \cite{AKNS}
\begin{equation}
\begin{bmatrix}
\phi(t,\lambda) &\bar{\phi}(t,\lambda)
\end{bmatrix}= \begin{bmatrix}
\bar{\psi}(t,\lambda) &\psi(t,\lambda)
\end{bmatrix}\mathcal{S}(\lambda).
\label{eqn:NFT_AKNS}
\end{equation}
The components $a(\lambda)$, $b(\lambda)$, $\bar{b}(\lambda)$ and $\bar{a}(\lambda)$ are known as the scattering data. The components $a(\lambda)$ and $b(\lambda)$ are sufficient to describe the signal completely. Their evolution along the $x$ dimension (along the length of the fiber) is simple \cite[Section III]{AKNS}
\begin{equation}
\begin{aligned}
a(x,\lambda)&=a(0,\lambda),\\
b(x,\lambda)&=b(0,\lambda)e^{-4i\lambda^2 x}.
\end{aligned}
\label{eqn:space_evolution}
\end{equation}
The \textit{reflection coefficient} is then defined as $\rho(\lambda)=b(\lambda)/a(\lambda)$ for $\lambda\in\mathbb{R}$ and it represents the continuous spectrum. In the case of $\kappa=1$, the nonlinear Fourier spectrum can also contain a so-called \textit{discrete spectrum}. It corresponds to the complex poles of the reflection coefficient in the upper half-plane $\mathbb{H}$, or equivalently to the zeros $\lambda_k\in\mathbb{H}$ of $a(\lambda)$. It is known that there are only finitely many $(N)$ such poles. The poles $\lambda_k$ are also referred to as eigenvalues and a corresponding set of values $\rho_k:=b(\lambda_k)/\left.\frac{\mathrm{d} a(\lambda)}{\mathrm{d} \lambda}\right\rvert_{\lambda=\lambda_k}$ are known as residues \cite[App. 5]{AKNS}. There are different ways to define a nonlinear Fourier spectrum. One possibility is $\{\rho(\lambda)\}_{\lambda\in\mathbb{R}}$, $(\lambda_k, \rho_k)_{k=1}^N$ \cite{AKNS}. The other is $\{b(\lambda)\}_{\lambda\in\mathbb{R}}$, $(\lambda_k, b(\lambda_k))_{k=1}^N$ \cite{Faddeev2007}. In this paper we are primarily interested in computation of $\rho(\lambda)$ but some notes regarding computation of $b(\lambda)$ and the $\lambda_k$ will also be given. Although we will illustrate our algorithms by applying them to the specific case of NFT of NSE with vanishing boundary condition, it should be noted that we in fact presenting algorithms for solving a class of equations similar to \eqref{eqn:ZS} \cite[Eq. 2]{AKNS}. Hence the algorithms presented in this paper should carry over to NFTs w.r.t. other nonlinear evolution equations such as the Korteweg--de Vries equation \cite{splittings} and other boundary conditions. 
\section{Numerical Computation of NFT using Higher Order Exponential Integrators}\label{NFT}
In this section we will start by outlining some assumptions that are required for the numerical methods that will be presented. We will give a brief overview of one of the approaches for computing the NFT and then talk specifically about implementations using commutator-free exponential integrators. 
To evaluate the methods, we describe examples and performance criteria. We will finally show and compare the results for various methods applied to the mentioned examples.
   
\textbf{We remark that only one of the investigated commutator-free exponential integrators can later serve as basis for our new fast method. However, the remaining higher order integrators have their own merits when the number of samples is low, since (asymptotically) slow NFT algorithms can be faster than (asymptotically) fast NFT algorithms in that regime.}

\subsection{Assumptions}\label{assumption}
Just like the FFT, the numerical computation of the NFT is carried out with finitely many discrete data samples. Hence, we need to make the following assumptions:
\begin{enumerate}
	\item The support of the signal $q(t)$ is truncated to a finite interval, $t\in[T_-,T_+]$, instead of $t\in(-\infty,\infty)$. The values $T_\pm$ are chosen such that the resulting truncation error is sufficiently small. The approximation is exact if $q(t) = 0\  \forall \  t\notin[T_-,T_+]$.
	\item The interval $[T_-,T_+]$ is divided into $D$ subintervals of width $h=(T_+-T_-)/D$. We assume that the signal is sampled at the midpoints of each subinterval $t_n=T_-+(n+0.5)h,\quad n=0,1,\ldots,D-1$ such that $q_n:=q(t_n)$.
\end{enumerate}
\subsection{Numerical Scattering}
The main step in numerically computing the NFT is to solve the scattering problem \eqref{eqn:ZS} for $\phi(T_+,\lambda)$ for different values of $\lambda$. We can view the $D$ subintervals as layers which scatter the eigenfunction $\phi(t,\lambda)$ as it moves from $t=T_-$ to $t=T_+$.  Using numerical ODE solvers we solve for an approximation $\hat{\phi}(T_+,\lambda)$ of $\phi(T_+,\lambda)$. By taking $\bar{\psi}(T_+,\lambda)$ and $\psi(T_+,\lambda)$ equal to the limit in \eqref{eqn:Jost} at $t=T_+$, we can compute with \eqref{eqn:NFT_AKNS} a numerical approximation of the scattering data and ultimately the reflection coefficient.

\subsection{Exponential Integrators}\label{sec:CFQM}
Almost any numerical method available in literature for solving ODEs can be used to solve for $\phi(T_+,\lambda)$\cite{RK4,Survey_Aston}. However, we are particularly interested in so-called exponential type integrators. These methods have been shown to provide a very good trade-off between accuracy and computational cost in several numerical benchmark problems while being fairly easy to implement, see \cite{Blanes2009} and references therein. We propose to use a special sub-class known as commutator-free quasi-Magnus (CFQM) exponential integrators as some NFT algorithms based on these integrators turn out to have the special structure \cite{PF} that is needed to make them fast. We show this in Section \ref{FNFT}.\\
The results in \cite{CFI} provide a scheme to compute a numerical approximation $\hat{\phi}(T_+,\lambda)$ of $\phi(T_+,\lambda)$. We start by fixing $\hat{\phi}(T_+,\lambda)=H(\lambda)\phi(T_-,\lambda)$, where
\begin{equation}
\begin{aligned}
H(\lambda) &= \left(\prod_{n=0}^{D-1}G_n(\lambda)\right)\\
&= G_{D-1}(\lambda)\, G_{D-2}(\lambda)\cdots G_0(\lambda),\\
\end{aligned}
\label{eqn:Tmatrix}
\end{equation}
with $n$ being the index of samples of $q(t)$.\\
The structure of $G_n(\lambda)$ depends on the integrator and the exact values depend on the signal samples $q_n$ and the value of $\lambda$. For the integrator in \cite{CFI}, $G_n(\lambda)=\text{CF}_J^{[r]}(t_n,\lambda)$ which leads to the following iterative scheme:
\begin{equation}
\begin{aligned}
\hat{\phi}_{n+1}(\lambda)&=\text{CF}_J^{[r]}(t_n,\lambda)\phi_n(\lambda)\\
&=\prod_{j=1}^{J}\expm({B_j(t_n,\lambda)})\phi(t_n,\lambda)\\
&=\expm({B_J(t_n,\lambda)})\cdots \expm({B_1(t_n,\lambda)})\phi_n(\lambda)\\
&=\phi(t_{n+1},\lambda)+\mathcal{O}(h^{r+1}),
\end{aligned}
\label{eqn:CF}
\end{equation}
where $\expm$ is the matrix exponential
\begin{equation}
\begin{aligned}
B_j(t_n,\lambda)&=h\sum_{k=1}^{K}a_{jk}C_k(t_n,\lambda),\quad j\in\{1,\ldots,J\},\\
C_k(t_n,\lambda)&=C(t_n+(c_k-0.5)h,\lambda),
\end{aligned}
\label{eqn:B_j}
\end{equation}
where $a_{jk}$ and $ c_k\in[0,1]$ for $k\in \{1,\ldots,K\}$ are constants that are specific to the integrator and $C(t,\lambda)$ as in \eqref{eqn:ZS}. By iterating with \eqref{eqn:Tmatrix} from $n=0,1,\ldots,D-1$, we obtain the numerical approximation of $\phi(T_+, \lambda)$ that we need to compute the NFT. 

For an integrator $\text{CF}_J^{[r]}$, $r$ is the order and $J$ is the number of matrix exponentials required for each subinterval. $K$ is the number of points within each subinterval where the signal value needs to be known. We refer the reader to \cite{CFI} for their derivation. 

An integrator of order $r$ has a local error (error in each subinterval) of $\mathcal{O}(h^{r+1})$. Over $D$ ($\propto1/h$) such subintervals i.e., over the interval $[T_-,T_+]$, the global error will be $\mathcal{O}(h^{r})$. This distinction of local and global error will become important when we define the error metric used to compare the various algorithms in Section \ref{sec:Numerical_examples}.\\
The integrator $\text{CF}_1^{[2]}$ is also sometimes referred to as the exponential midpoint rule. It was used in the context of NFT for the defocusing NSE ($\kappa=-1$) by Yamada and Sakuda \cite{YS} and later by Boffetta and Osborne\cite{BO}. For $\text{CF}_1^{[2]}$, \eqref{eqn:CF} reduces to
\begin{equation}
\begin{aligned}
\hat{\phi}_{n+1}(\lambda)&=G_n(\lambda)\phi_n(\lambda), \quad \text{where}\\
G_n(\lambda)&=\expm(hC_n(\lambda)),\\
C_n(\lambda)&=\begin{bmatrix}
-i\lambda &q_n\\-\kappa q_n^* &i\lambda
\end{bmatrix}.
\end{aligned}
\end{equation}
This is applied repeatedly as in \eqref{eqn:Tmatrix} to obtain $\hat{\phi}(T_+,\lambda)$.
In \cite{CF4_APC} we investigated the possibility of using $\text{CF}_2^{[4]}$ (first introduced in \cite{CF4_2}) to obtain $\hat{\phi}(T_+,\lambda)$. We were able to show its advantage over $\text{CF}_1^{[2]}$ when considering the trade-off between an error and execution time. Here we investigate further in this direction and evaluate $\text{CF}_3^{[4]}$, $\text{CF}_3^{[5]}$ and $\text{CF}_4^{[6]}$, which are fourth-, fifth- and sixth-order methods respectively.

The CFQM exponential integrators require multiple non-equispaced points within each subinterval. However, it is unrealistic to assume that signal samples at such non-equispaced points can be obtained in a practical setting. In \cite{CF4_APC} we used local cubic-spline based interpolation to obtain the non-equispaced points from the mid-points of each subinterval. (We will refer to the samples at these midpoints as the given samples.) However we found that local cubic-spline based interpolation is not accurate enough for higher-order methods. Here, we propose to utilize the Fourier transform and its time-shift property for interpolation, i.e.,
\begin{equation}
q(t-t_s)=\mathcal{F}^{-1}(e^{-i \xi t_s}\mathcal{F}(q(t))),
\label{eqn:Interpolation}
\end{equation} 
to obtain the samples on shifted time grids required for \eqref{eqn:B_j} with $t_s = -(c_k-0.5)h$. This interpolation rule is also known in signal-processing literature as sinc or bandlimited interpolation \cite[Section 7.4.2]{Oppenheim} and it is accurate when $q(t)$ is sampled in accordance with the Nyquist criterion. As we are working with discrete signal samples, the interpolation can be implemented efficiently using the FFT. The MATLAB code that we used can be found in Appendix B. We remark that we use band-limited interpolation for all methods that require non-equispaced samples: $\text{CF}_2^{[4]}$, $\text{CF}_3^{[4]}$, $\text{CF}_3^{[5]}$ and $\text{CF}_4^{[6]}$.
\subsection{Error Metric and Numerical Examples}\label{sec:Numerical_examples}
In this subsection, we compare the performance of CFQM exponential integrators  $\text{CF}_1^{[2]}$, $\text{CF}_2^{[4]}$, $\text{CF}_3^{[4]}$, $\text{CF}_3^{[5]}$ and $\text{CF}_4^{[6]}$, the two-step Implicit-Adams method (IA$_2$) introduced in \cite{IA2} and the fourth-order Runge-Kutta method \cite{RK4} (RK$_4$) for computation of the reflection coefficient. The fourth-order Runge-Kutta method ($r=4$) was the first fourth-order method used for the computation of the reflection coefficient in \cite{RK4,IT_NFT2}. We include the third-order two-step Implicit-Adams method ($r=3$) here as it was the first higher-order method that was introduced in the context of fast nonlinear Fourier transform. The meaning of "fast" will be made precise in Section \ref{FNFT}. \\
We are interested in evaluating the trade-off between the increased accuracy and execution time due to use of higher-order methods. We assess the accuracy of different methods using the relative $L^2$-error
\begin{equation}
E_{\rho}=\frac{\sqrt{\sum_{n=0}^{M-1}\lvert\rho(\lambda_n)-\hat{\rho}(\lambda_n)\rvert^2}}{\sqrt{\sum_{n=0}^{M-1}\lvert\rho(\lambda_n)\rvert^2}},
\label{eqn:Error_criterion}
\end{equation}
where $\rho(\lambda)$ is the analytical reflection coefficient, $\hat{\rho}(\lambda)$ is the numerically computed reflection coefficient and $\lambda_n$ are $M$ equally-spaced points in $[-\lambda_{\text{max}},\lambda_{\text{max}}]$. $E_{\rho}$ is a global error and hence it is expected to be $\mathcal{O}(h^{r})$ for an integrator of order $r$ as explained in Section \ref{sec:CFQM}. We compute the reflection coefficient at the same number of points $M$ as the number of given samples $D$, i.e. $M=D$, unless mentioned explicitly otherwise.

\subsubsection{Example 1: Hyperbolic Secant, $\kappa=1$}\label{sec:Example_1}
As the first numerical example we chose the signal $q(t)=\mathring{q}e^{-2i\lambda_0 t} \sech(t)$. The closed form of the reflection coefficient is given by applying the frequency-shift property \cite[Section D]{IT_NFT} to the analytical known reflection coefficient of the secant-hyperbolic signal \cite{sech},
\begin{equation}
\begin{aligned}
\rho(\lambda)&=\frac{b(\lambda)}{a(\lambda)},\\
b(\lambda)&=\frac{-\sin(\pi)}{\cosh(\pi(\lambda-\lambda_0))},\\
a(\lambda)&=\frac{\Gamma^2(0.5-i\lambda)}{\Gamma(0.5-i(\lambda-\lambda_0)+\mathring{q})\Gamma(0.5-i(\lambda-\lambda_0)-\mathring{q})},
\end{aligned}
\end{equation} 
where $\Gamma(\cdot)$ is the gamma function. 
The discrete spectrum is
\begin{align}
\lambda_k &= \lambda_0+i(\mathring{q}+0.5-k), \quad k=1,2,\ldots,M_D,\\
b_k &= (-1)^{k},\quad k=1,2,\ldots,M_D,\\
M_D &= \floor{(\mathring{q}+0.5)}.
\end{align}
We set $\mathring{q}=5.4$, $\lambda_0 = 3$, $\lambda_{\text{max}}=10$, and chose $[T_-,T_+]=[-32,32]$ to ensure negligible truncation error.

\subsubsection{Example 2: Rational Reflection Coefficient with one pole, $\kappa=1$}\label{sec:Example_2}
The signal is given by \cite{Rational_focusing}
\begin{equation}
q(t) = \begin{cases}
-2i\gamma\frac{\alpha}{\lvert\alpha\rvert}\sech\Big(2\gamma t+\text{arctanh}\Big(\frac{\beta}{\gamma}\Big)\Big), &t\leq0\\
0,&t>0,
\end{cases}
\label{eqn:Rat_rho_signal}
\end{equation}
where $\alpha$, $\beta$ are scalar parameters and $\gamma=\sqrt{\alpha \alpha^*+\beta^2}$. We used $\alpha=1$ and $\beta=-1$. The corresponding reflection coefficient is then known to be 
\begin{equation}
\rho(\lambda)=\frac{\alpha}{\lambda-i\beta}.
\end{equation}
We set $\lambda_{\text{max}}=60$ and chose $[T_-,T_+]=[-12,0]$. As the signal in \eqref{eqn:Rat_rho_signal} has a discontinuity, it cannot be interpolated well using bandlimited interpolation. We hence assume only in this example that we can sample the signal at exactly the points that we require. 

\subsubsection{Example 3: Hyperbolic Secant, $\kappa=-1$}\label{sec:Example_3}
The signal is given by
\begin{equation}
q(t) = \frac{\mathcal{Q}}{L} \Bigg(\sech\bigg(\frac{t}{L}\bigg)\Bigg)^{1-2i\mathcal{G}},
\end{equation}
where $\mathcal{G}$, $\mathcal{L}$ and $\mathcal{Q}$ are scalar parameters. We used $\mathcal{G}=1.5$, $\mathcal{L}=0.04$ and $\mathcal{Q}=5.5$. The corresponding reflection coefficient is known to be \cite{sech_defocusing}
\begin{equation}
\rho(\lambda)=-2^{-2i\mathcal{G}}\mathcal{Q}\frac{\Gamma(d)\Gamma(f_-)\Gamma(f_+)}{\Gamma(d^*)\Gamma(g_-)\Gamma(g_+)},
\end{equation}
where $\Gamma(\cdot)$ is the gamma function, $d=0.5+i(\lambda\mathcal{L}-\mathcal{G})$, $f_\pm = 0.5-i(\lambda\mathcal{L}\pm \sqrt{\mathcal{G}^2+\mathcal{Q}^2})$, and $g_\pm = 1-i(\mathcal{G}\pm \sqrt{\mathcal{G}^2+\mathcal{Q}^2})$. We set $\lambda_{\text{max}}=250$ and chose $[T_-,T_+]=[-1.5,1.5]$.\\

\Figure[t][width=3in,height=2.5in]{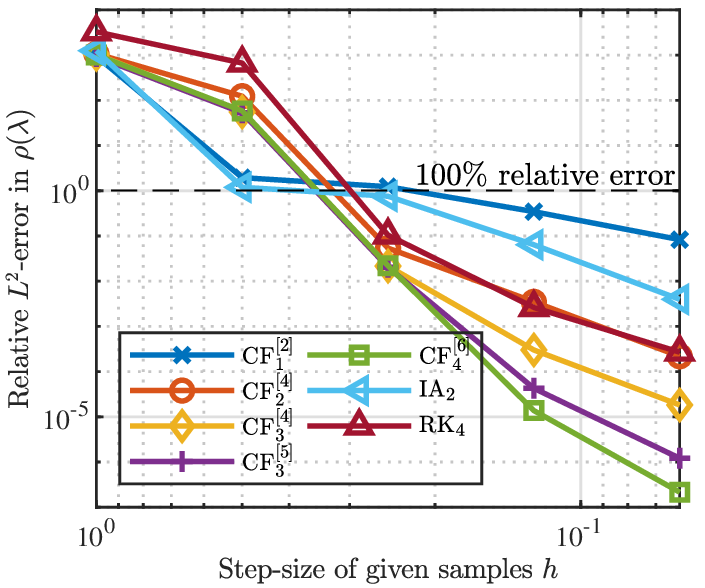}
{Error using slow NFT algorithms for Example 1 with $\lambda_{\text{max}}=1$.
	\label{fig:CF_sech_focusing_low_error} }

\Figure[t][width=3in,height=2.5in]{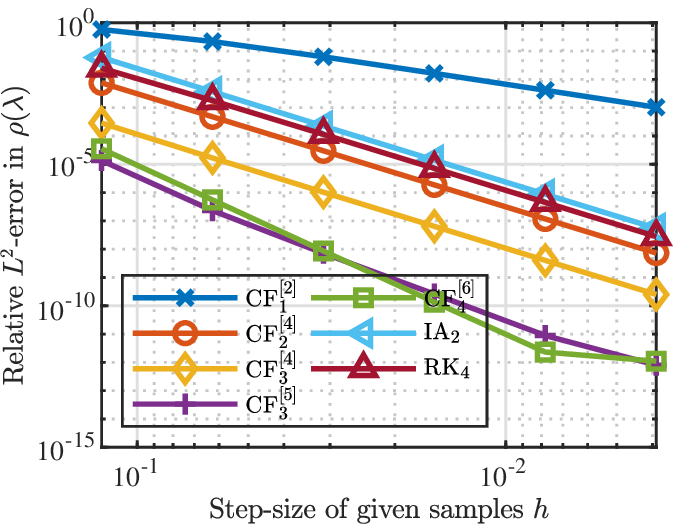}
{Error using slow NFT algorithms for Example 1 with $\lambda_{\text{max}}=10$. \label{fig:CF_sech_focusing_error}}

\Figure[t][width=3in,height=2.5in]{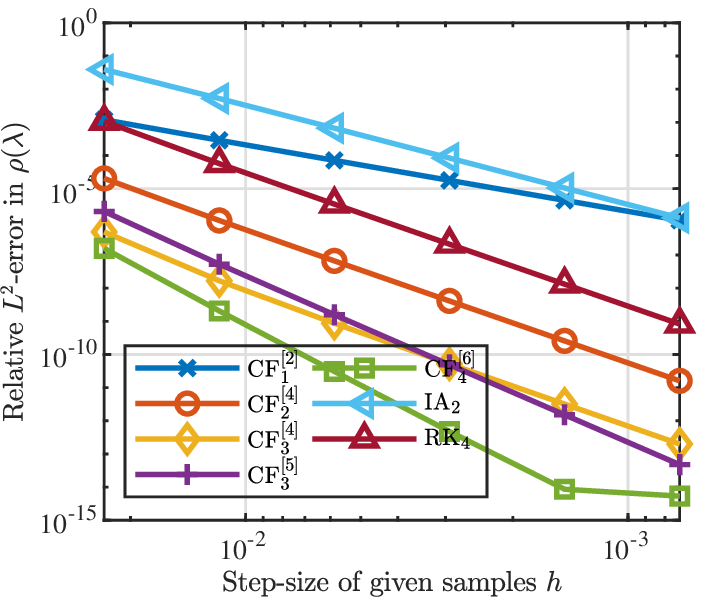}
{Error using slow NFT algorithms for Example 2.
\label{fig:CF_rational_1_pole_error}}

\Figure[t][width=3in,height=2.5in]{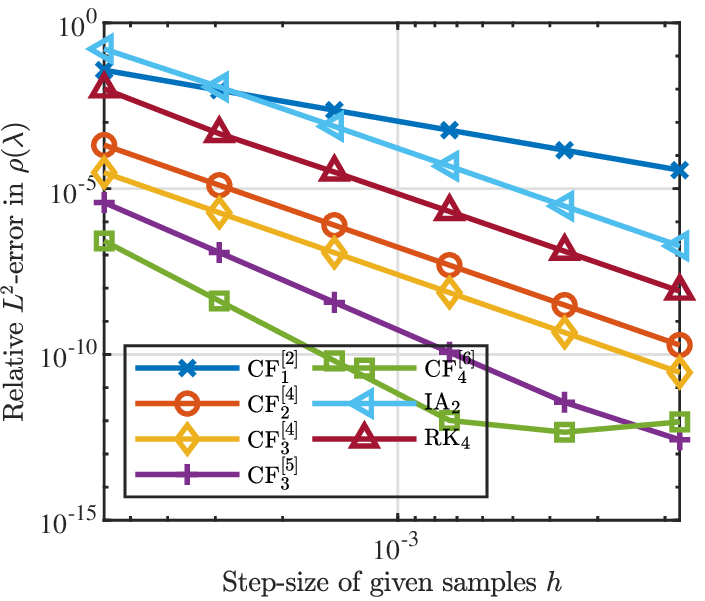}
{Error using slow NFT algorithms for Example 3.
\label{fig:CF_sech_defocusing_error}}

The numerical methods were implemented and tested in 64-bit MATLAB (R2018a) running in Ubuntu 16.04 on a machine with an Intel\textsuperscript{\textregistered} Core\texttrademark{} i7-5600U CPU with a maximum clock rate of 3200 MHz and 8192 MB of DDR3 memory at 1600 MHz. The CPU was set to the highest available performance setting and the number of computational threads was set to 1 using the \texttt{maxNumCompThreads} function of MATLAB. The closed-form expression of a $2\times 2$ matrix exponential as in \cite{Explicit_expm} was used for the CFQM exponential integrators. 

As we are interested in studying the complexity-accuracy trade-off of the NFT algorithms, we need a measure of computational complexity. In the literature, either number of floating  point operations (FLOPs) or execution times are used as a measure of the computational complexity. Both are not ideal. Although FLOP counting seems more objective, in practice FLOP counts are (just like execution times) implementation specific and it is hard to determine even the number of FLOP counts of basic operations such as square roots. FLOP counts also do not account for typical capabilities of modern processors and neglect critical issues such as memory access. We will present our results in terms of execution times as we believe that they are more realistic than FLOP counts. However, to ensure that our implementations were equally efficient, we carried out an additional FLOP count analysis in Appendix A. By comparing the FLOP counts with the measured execution times we show there that the measured execution times agree well with the FLOP counts for medium to high number of samples. We also show there that the FLOP counts are not representative of computation costs for low number of samples. 

Execution times were recorded using the MATLAB stopwatch function (tic-toc). We report the best execution time among three runs to ensure that we minimize the impact of unrelated background processes.    

Fig. \ref{fig:CF_sech_focusing_low_error} shows the error measure $E_{\rho}$, as defined in \eqref{eqn:Error_criterion}, for Example 1 for a range of relatively large step-sizes $h$. To read such error plots we look at the error achieved by each method for a particular step-size. For the largest two step-sizes, all the errors are above 100 percent and hence a comparison of the methods is not meaningful. The remaining results suggest that the higher-order methods can always be preferred over the lower-order methods. 

The error measure $E_{\rho}$ for smaller sampling intervals $h$ for Example 1, 2, and 3 are shown in Figs. \ref{fig:CF_sech_focusing_error}, \ref{fig:CF_rational_1_pole_error} and \ref{fig:CF_sech_defocusing_error} respectively.\footnote{To ensure that the discontinuity in Example 2 is faithfully captured, we use $t_n=T_-+nh$ for the Runge-Kutta method and the Implicit-Adams method, instead of the description in Section \ref{assumption}.} For all three examples, the slopes of the error-lines are in agreement with the order $r$ of each method except for IA$_2$. For smooth signals IA$_2$ is seen to have an error of order four rather than the expected three. This observation is in agreement with \cite[Fig. 2]{IA2}. However, for the discontinuous signal of Example 2 we see third-order behavior as expected. We can also see that a higher $r$ generally corresponds to better accuracy (lower $E_{\rho}$) for the same $h$. However, that is not necessarily obvious as seen in Fig. \ref{fig:CF_sech_focusing_error}, where $\text{CF}_3^{[5]}$ is more accurate than $\text{CF}_4^{[6]}$ for larger $h$. The advantage of using three exponentials ($J=3$) in $\text{CF}_4^{[3]}$ instead of two in $\text{CF}_4^{[2]}$ is also clear from the figures. The third-order Implicit-Adams method (IA$_2$ with $r=3$) and fourth-order Runge-Kutta method (RK$_4$) may be more accurate than $\text{CF}_2^{[1]}$ depending on the signal and other parameters, but have lower accuracy compared to $\text{CF}_4^{[2]}$ and $\text{CF}_4^{[3]}$. 

The error $E_{\rho}$ reaches a minimum around $10^{-12}$ and can start rising again as seen in Fig. \ref{fig:CF_sech_defocusing_error} for $\text{CF}_6^{[4]}$. To understand this behavior, note that the local error in \eqref{eqn:CF}  is actually $\mathcal{O}(h^{r+1}+\varepsilon)$, where $\varepsilon$ is a small constant due to finite precision effects that can normally be neglected. The global error is thus $\mathcal{O}(h^r+\varepsilon h^{-1})$. As $h$ is becoming smaller and smaller, the second component also known as the arithmetic error, becomes dominant and eventually causes the total error to rise again \cite{Liu2006}.

For the CFQM exponential integrators, computation of the transfer-matrix $H(\lambda)$ in \eqref{eqn:Tmatrix} for each $\lambda$ requires $JD$ multiplications of $2\times 2$ matrices \eqref{eqn:CF} for $D (\propto 1/h)$ given samples. If the reflection coefficient is to be computed at $D$ points then the overall computational complexity will be of the order $\mathcal{O}(D^2)$. In Fig. \ref{fig:CF_sech_focusing_time} we plot the execution times of all the methods for Example 1. These execution times are representative for all examples. We can see that the execution time scales quadratically with $1/h$. The execution time of the CFQM exponential integrators is approximately a linear function of $J$. The IA$_2$, RK$_4$ and $\text{CF}_4^{[2]}$ methods have similar execution times. Although both $\text{CF}_4^{[3]}$ and $\text{CF}_5^{[3]}$ methods require 3 matrix exponentials, the execution times of $\text{CF}_5^{[3]}$ are higher because it involves more operations using complex numbers compared to $\text{CF}_4^{[3]}$. 

To evaluate the trade-off between the execution time and accuracy, we plot the execution time on the x-axis and the error on the y-axis in Fig. \ref{fig:CF_sech_focusing_tradeoff} for Example 1. To read such trade-off plots we look at the error achieved by each method for a given amount of time. For Example 1 it turns out that $\text{CF}_5^{[3]}$ provides the best trade-off, but we can conclude that extra computation cost of the higher-order methods is generally justified by increased accuracy. 

Although performing matrix multiplications of $2\times 2$ matrices is fast, the total cost of the NFT ($\mathcal{O}(D^2)$) is significantly higher when compared to its linear analogue, the FFT, which has a complexity of only $\mathcal{O}(D\log D)$. So the natural question to ask is: Can the complexity be reduced? -- Yes, this will be shown in the next section.

\Figure[t][width=3in,height=2.5in]{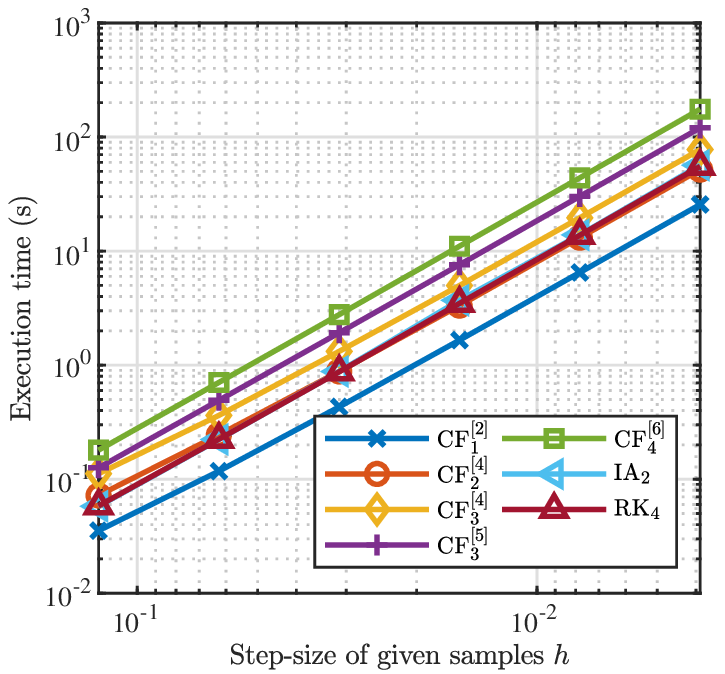}
{Execution time using slow NFT algorithms for Example 1.
\label{fig:CF_sech_focusing_time}}

\Figure[t][width=3in,height=2.5in]{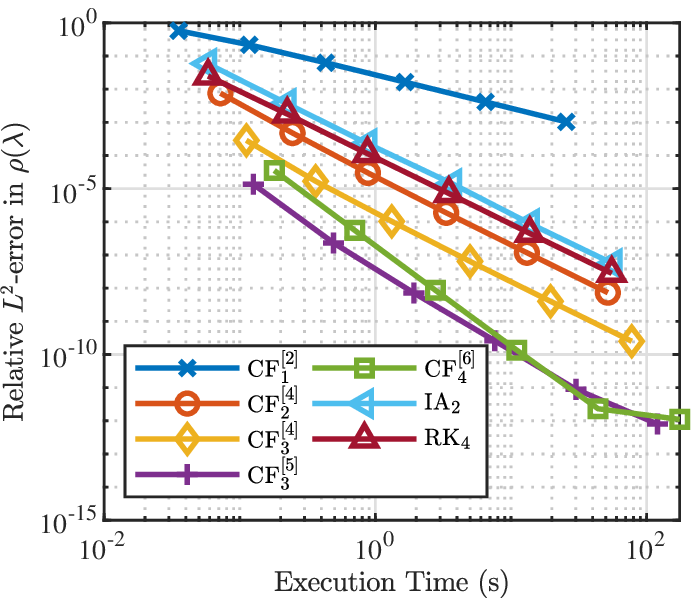}
{Error vs. Execution time trade-off using slow NFT algorithms for Example 1.
\label{fig:CF_sech_focusing_tradeoff}}

\section{Fast Fourth-Order NFT}\label{FNFT}
In this section we investigate which of the new higher-order NFT algorithms from the previous section can be made fast by using suitable splittings of the matrix exponential. \textbf{We find that only the $\text{CF}_4^{[2]}$ NFT can be made fast. The result is a fast fourth-order NFT algorithm. We then compare the slow $\text{CF}_4^{[2]}$ NFT with its fast variant to ensure that the gain in computational complexity outweighs the loss in accuracy introduced by approximations of the  matrix exponential.}

\subsection{Fast Scattering Framework}\label{fast_scattering}
In the framework proposed in \cite{PF}, each matrix $G_n(\lambda)$ is approximated by a rational function matrix $\hat{G}_n(z)$, where $z=z(\lambda)$ is a transformed coordinate. By substituting these approximations in \eqref{eqn:Tmatrix}, a rational function approximation $\hat{H}(z)$ of $H(\lambda)$ is obtained.
\begin{equation}
\hat{H}(z)=\prod_{n=0}^{D-1}\hat{G}_n(z).
\label{eqn:H_approx}
\end{equation}
We want to compute the coefficients of the numerator and denominator polynomials, respectively. A straightforward implementation of the matrix multiplication where each entry is a polynomial, has a complexity of $\mathcal{O}(D^2)$. Instead, by using a binary-tree structure and FFTs \cite[Alg. 1]{PF}, the computational complexity can be reduced to $\mathcal{O}(D\log^2D)$. Hence it is referred to as fast scattering. In \cite{PF}, the number of samples $D$ was assumed to be a power of two. In cases where $D$ is not a power of two, we use the following approach. We write $D = 2^{D_1}+2^{D_2}+\ldots +2^{D_m}$, where $D_1$, $D_2,\ldots$$, D_m$ are non-negative integers. We first choose $D_1$ as large as possible. Then we choose $D_2$ as large as possible and repeat until all $D_k$ are fixed. This step splits the $D$ samples into $m$ sets to each of which the fast scattering is applied. The results $\hat{H}_1(z), \hat{H}_2(z) \ldots \hat{H}_m(z)$  are then multiplied using the rule $\hat{H}(z)=[\ldots[[\hat{H}_1(z)\hat{H}_2(z)]\hat{H}_3(z)]\ldots]\hat{H}_m(z)$. Each multiplication is carried out using the same FFT based algorithm as in \cite{PF}. 

The rational function approximation $\hat{H}(z)$ is explicitly parametrized in $z$ and hence \eqref{eqn:Tmatrix} is reduced to polynomial evaluations for each $z$. To elaborate, we again restrict ourselves to $G_n(\lambda)$ of the form \eqref{eqn:CF}. Hence for $\text{CF}_1^{[2]}$, we need to approximate $G_n(\lambda) = \expm(hC_n(\lambda))$. The matrix exponential can be approximated to various orders of accuracy using rationals \cite{RatApprox} or using splitting-schemes such as the well-known Strang-splitting and the higher-order splitting-schemes developed in \cite{splittings}. The splitting-schemes map $\lambda \in \mathbb{R}$, the domain of the reflection coefficient, to $z = \text{exp}(i\lambda h/m)$ on the unit circle, where $m$ is a real rational. Such mappings have a certain advantage when it comes to polynomial evaluations which we cover in Section \ref{sec:fast_eval}. Note that the mapping $z = e^{i\lambda h/m}$ is periodic in $\lambda$ with period $2\pi m/h$. Hence we can resolve the range $\lvert \text{Re}(\lambda)\rvert<\pi m/h$. (See e.g. \cite{Skaar_gratings}.) This is similar to the Nyquist--Shannon sampling theorem for the FFT.

For a higher-order $\text{CF}_J^{[r]}$ integrator, each $G_n(\lambda)$ is a product of $J$ matrix exponentials. For example let us look at $\text{CF}_2^{[4]}$. We can write
\begin{equation}
\begin{aligned}
G_n(\lambda)&\!=\!\expm(hC^2_n(\lambda))\, \expm(hC^1_n(\lambda)),\\
C^2_n(\lambda)&\!=\!a_2C(T_-\!+\!(n\!+\!c_1)h,\lambda)\!+\!a_1C(T_-\!+\!(n\!+\!c_2)h,\lambda),\\
C^1_n(\lambda)&\!=\!a_1C(T_-\!+\!(n\!+\!c_1)h,\lambda)\!+\!a_2C(T_-\!+\!(n\!+\!c_2)h,\lambda),\\
a_1 &= \frac{1}{4}+\frac{\sqrt{3}}{6},\quad a_2 = \frac{1}{4}-\frac{\sqrt{3}}{6},\\
c_1 &= \frac{1}{2}-\frac{\sqrt{3}}{6}, \quad c_2 = \frac{1}{2}+\frac{\sqrt{3}}{6}.
\end{aligned}
\end{equation}

Each of the two matrix exponentials can be approximated individually using a splitting-scheme from \cite{splittings}. $\hat{H}(z)$ can then be obtained as in \eqref{eqn:H_approx}. However, there are a few caveats which prevent extension of this idea to higher-order methods. The splitting-schemes in \cite{splittings} should not be applied to CFQM exponential integrators with complex coefficients $a_{jk}$. Complex coefficients mean that $\lambda \in \mathbb{R}$ is no longer mapped to $z$ on the unit circle. Such a mapping is undesirable for polynomial evaluation as will be explained in Section \ref{sec:fast_eval}. In addition, we do not even obtain a polynomial structure if there exists no $z$ such that $\text{exp}(i\lambda h\sum_{k=1}^{K}a_{j,k})$ is an integer power of this $z$ for all $j$. Furthermore, if such a $z$ exists but only for high co-prime integer powers, $\hat{G}_n(\lambda)$ will consist of sparse polynomials of high degree, which can significantly hamper the computational advantage of using the approximation. Due to these reasons we restrict ourselves to fast implementations of $\text{CF}_1^{[2]}$ and $\text{CF}_2^{[4]}$ which will be referred to as $\text{FCF}_1^{[2]}$ and $\text{FCF}_2^{[4]}$. Even though we made the $\text{FCF}_1^{[2]}$ algorithm available in the FNFT software library \cite{software} already, accuracy and execution times for it haven't been assessed and published formally anywhere in literature. The $\text{FCF}_2^{[4]}$ algorithm is completely new. For both $\text{FCF}_1^{[2]}$ and $\text{FCF}_2^{[4]}$ we use the fourth-order accurate splitting \cite[Eq. 20]{splittings}.
\subsection{Fast Evaluation}\label{sec:fast_eval}
Once we obtain the rational function approximation $\hat{H}(z)$ of $H(\lambda)$ in terms of numerator and denominator coefficients, we only have to evaluate the numerator and denominator polynomials for each value of $z=z(\lambda)$ in order to compute the reflection coefficient. The degree of the polynomials to be evaluated will be at least $D$ which can be in the range of $10^3$--$10^4$. It is known that evaluation of such high-degree polynomials for large values of $z$ can be numerically problematic \cite[Section IV-E]{FNFT}. However, by choosing the mapping $z = \text{exp}(i\lambda h/m)$, which maps the real line to the unit circle, the polynomials need to be evaluated on the unit circle where evaluation of even high-degree polynomials is numerically less problematic. The higher-order splitting schemes in \cite{splittings} were developed with such a mapping in mind allowing for approximations of the matrix exponentials as rational functions in $z$.  Evaluating any polynomial of degree $D$ using Horner's method has a complexity of $\mathcal{O}(D)$ \cite[Section IV-E]{FNFT}. Hence for $M$ values of $z$, the total cost of fast scattering followed by polynomial evaluation would be $\mathcal{O}(D\log^2D)+\mathcal{O}(MD)$. 

Mapping  $\lambda \in \mathbb{R}$ to $z$ on the unit circle has an additional computational advantage. Let $p(z)=p_Nz^N+p_{N-1}z^{N-1}+\ldots+p_0$ be a polynomial in $z$ of degree $N$. Evaluation of $p(z)$ at a point $z_k$ can be written as
\begin{equation}
p(z_k)=\sum_{n=0}^{N} p_n z_k^n
=z_k^N\sum_{n=0}^{N} p_{N-n} z_k^{-n}.
\label{eqn:fast_polyval}
\end{equation} 
For $M$ equispaced points $z_k,\ k=1,\ldots,M$, on a circular arc, this amounts to taking the chirp Z-transform (CZT) of the polynomial coefficients. The CZT can be computed efficiently using the algorithm in \cite{Rabiner1969} which utilizes FFTs. We can also see \eqref{eqn:fast_polyval} as a non-uniform discrete Fourier transform of the polynomial coefficients which allows us to utilize efficient non-uniform FFT (NFFT) algorithms in \cite{NFFT} for evaluating the polynomial. If the number of evaluation points $M$ is in the same order of magnitude as $D$, the complexity of evaluation becomes $\mathcal{O}(D\log D)$ and hence the overall complexity of the fast nonlinear Fourier transform (FNFT) is $\mathcal{O}(D\log^2D)$. In the next section we will see that the error of the $\text{FCF}_2^{[4]}$ algorithm reaches a minimum value and then starts rising. This is again due to the arithmetic error as we already saw in Section \ref{sec:Numerical_examples}. We remark that in numerical tests the CZT was found to perform equally well as the NFFT before the error minimum but the error rise thereafter was significantly steeper. We hence used the NFFT routine from \cite{NFFT} for evaluating the polynomials.
\subsection{Numerical Examples}
\subsubsection{Reflection coefficient}
We now compare the implementations of $\text{CF}_1^{[2]}$ and $\text{CF}_2^{[4]}$ presented in Section \ref{sec:Numerical_examples} and their fast versions $\text{FCF}_1^{[2]}$ and $\text{FCF}_2^{[4]}$ for computing the reflection coefficient $\rho(\lambda)$. We plot the error versus the execution time for Example 1 in Fig. \ref{fig:FCFvsCF_sech_focusing_tradeoff}, for Example 2 in Fig. \ref{fig:FCFvsCF_rational_1_pole_tradeoff} and for Example 3 in Fig. \ref{fig:FCFvsCF_sech_defocusing_tradeoff}. In the three figures we can see that the fast FCF algorithms achieve similar errors as their slow CF counterparts in a significantly shorter time. From the other viewpoint, for the same execution time, the FCF algorithms achieve significantly lower errors compared to CF algorithms. Our new algorithm $\text{FCF}_2^{[4]}$ outperforms $\text{FCF}_1^{[2]}$ in the trade-off for all the examples which again highlights the advantage of using higher-order CFQM exponential integrators. 

\Figure[t][width=3in,height=2.5in]{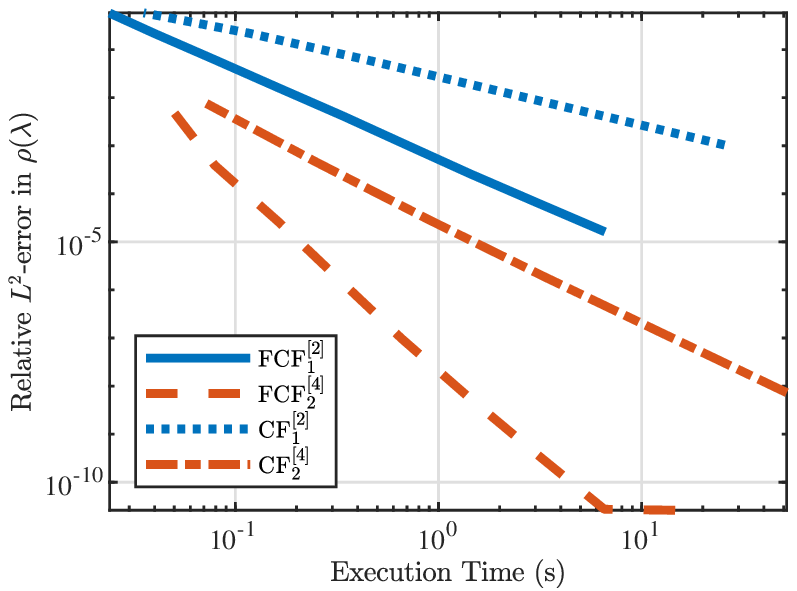}
{Error vs. Execution time trade-off using CF and FCF algorithms for Example 1.
\label{fig:FCFvsCF_sech_focusing_tradeoff}}

\Figure[t][width=3in,height=2.5in]{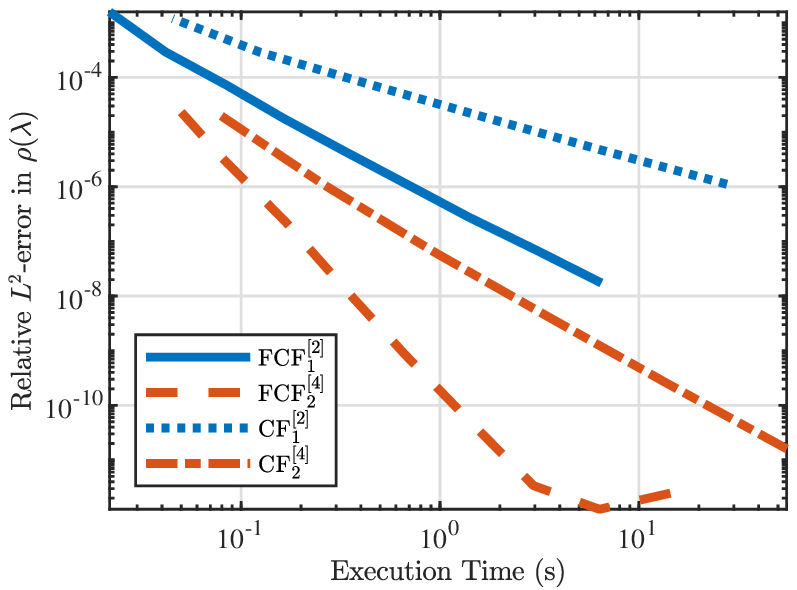}
{Error vs. Execution time trade-off using CF and FCF algorithms for Example 2.
\label{fig:FCFvsCF_rational_1_pole_tradeoff}}

\Figure[t][width=3in,height=2.5in]{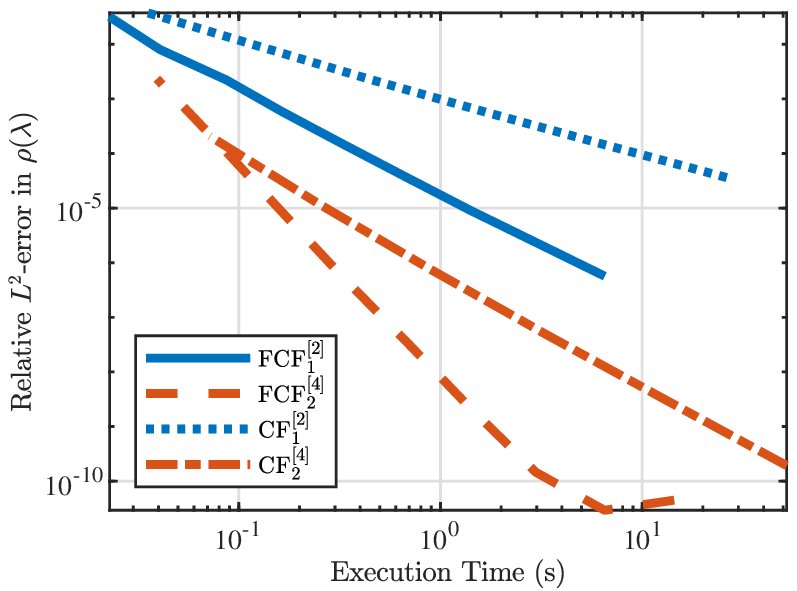}
{Error vs. Execution time trade-off using CF and FCF algorithms for Example 3.
\label{fig:FCFvsCF_sech_defocusing_tradeoff}}

Since the NFT is a nonlinear transform, it changes its form under signal amplification, and computing it typically becomes increasingly difficult when a signal is amplified \cite{Survey_Aston}. Hence it is of interest to study amplification of error with increase in signal amplitude. To test the amplification we use Example 1 and sweep the signal amplitude $\mathring{q}$ from $0.4$ to $10.4$ in steps of $1.0$ while keeping all other parameters the same as before. As the time-window remains the same, amplification the signal amplitude leads to directly proportional amplification of the signal energy. We compute the error $E_{\rho}$ for decreasing $h$ for each value of $\mathring{q}$ for the CF and FCF algorithms. We plot $E_{\rho}$ versus the sampling interval $h$ on a log-scale for CF algorithms in Fig. \ref{fig:CF_sech_focusing_EvsA} and for FCF algorithms in Fig. \ref{fig:FCF_sech_focusing_EvsA}. Instead of plotting individual lines for each value of $\mathring{q}$, we represent the amplitude using different shades of gray. As shown in the colourbar, lighter shades represents lower $\mathring{q}$ and darker shades represent higher $\mathring{q}$. The stripes with a higher slope are the higher-order methods. All the four algorithms i.e., $\text{CF}_1^{[2]}$, $\text{CF}_2^{[4]}$, $\text{FCF}_1^{[2]}$ and $\text{FCF}_2^{[4]}$ show similar trends for the amplification of error with signal amplitude. The $\text{CF}_1^{[2]}$ algorithm was compared with other methods in \cite{Survey_Aston} (where it is referred to as BO), and they conclude that $\text{CF}_1^{[2]}$ scales the best with increasing signal amplitude. Hence the results shown in Fig. \ref{fig:CF_sech_focusing_EvsA} are very motivating as the amplification in the error of $\text{CF}_2^{[4]}$ is similar to the amplification for $\text{CF}_1^{[2]}$. The error of approximations used in the FCF algorithms also depends on $\mathring{q}$. However, comparing Fig. \ref{fig:CF_sech_focusing_EvsA} and Fig. \ref{fig:FCF_sech_focusing_EvsA} we can see that the contribution of the approximation error is small. These results combined with the results in the trade-off plots (Figs. \ref{fig:FCFvsCF_sech_focusing_tradeoff}, \ref{fig:FCFvsCF_rational_1_pole_tradeoff}, and \ref{fig:FCFvsCF_sech_defocusing_tradeoff}) make a strong case for our new $\text{FCF}_2^{[4]}$ algorithm.\\
 
\Figure[t][width=3in,height=2.5in]{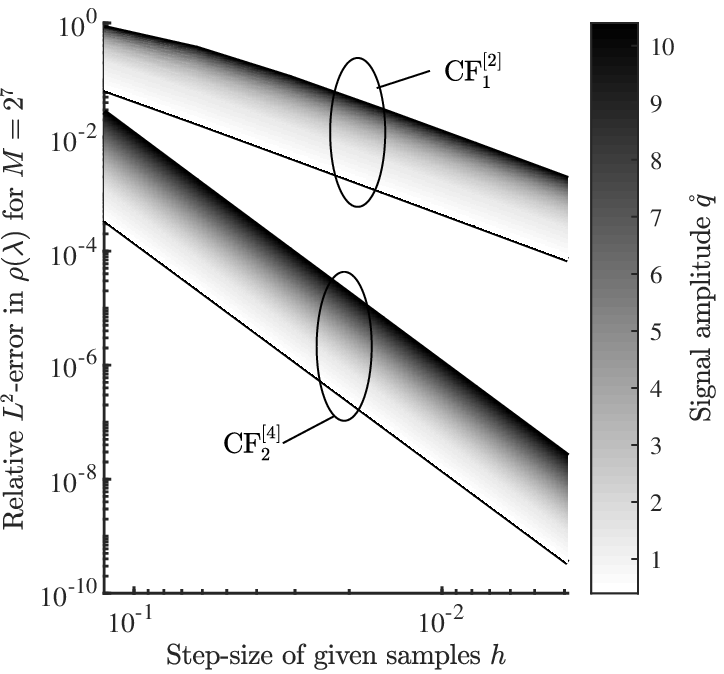}
{Variation of error of CF algorithms with amplitude for Example 1. The fourth-order $\text{CF}_2^{[4]}$ algorithm is seen to have gradual increase in error with increase in amplitude similar to the second-order $\text{CF}_1^{[2]}$ algorithm.
\label{fig:CF_sech_focusing_EvsA}}
 
\Figure[t][width=3in,height=2.5in]{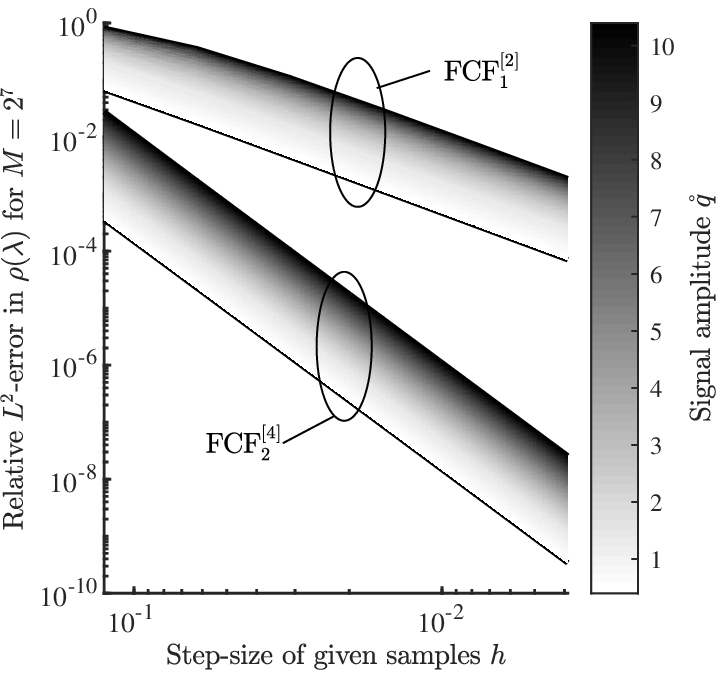}
{Variation of error of FCF algorithms with amplitude for Example 1. Approximating the matrix exponentials with splitting schemes does not significantly affect the amplification of error with increasing amplitude.
\label{fig:FCF_sech_focusing_EvsA}}
\subsubsection{b-coefficient}
The accurate and fast computation of the scattering coefficient $b(\lambda)$ (Section \ref{Preliminaries}) is of interest to the fiber-optic communication community, as an efficient FNFT algorithm can be combined with the recently proposed $b$-modulation \cite{bmod,bmod_new,Le2018} scheme to develop a complete NFT based fiber-optic communication system. Hence to test the performance of both the FCF algorithms in computation of the $b$-coefficient, we define
\begin{equation}
E_{b} =\frac{\sqrt{\sum_{n=0}^{M-1}\lvert b(\lambda_n)- \hat{b}(\lambda_n)\rvert^2}}{\sqrt{\sum_{n=0}^{M-1}\lvert b(\lambda_n)\rvert^2}},
\label{eqn:Error_criterion_b}
\end{equation}
where $b(\lambda)$ is the analytically known and $\hat{b}(\lambda)$ is the numerically computed scattering coefficient. For the numerical test we again use the signal from Example 1 as $b(\lambda)$ is known. We plot the error $E_b$ for both the FCF methods for decreasing sampling interval $h$ in Fig. \ref{fig:FCF_sech_focusing_error_b}. $\text{FCF}_2^{[4]}$ clearly outperforms $\text{FCF}_1^{[2]}$ even after considering the additional computational cost. 

\Figure[t][width=3in,height=2.5in]{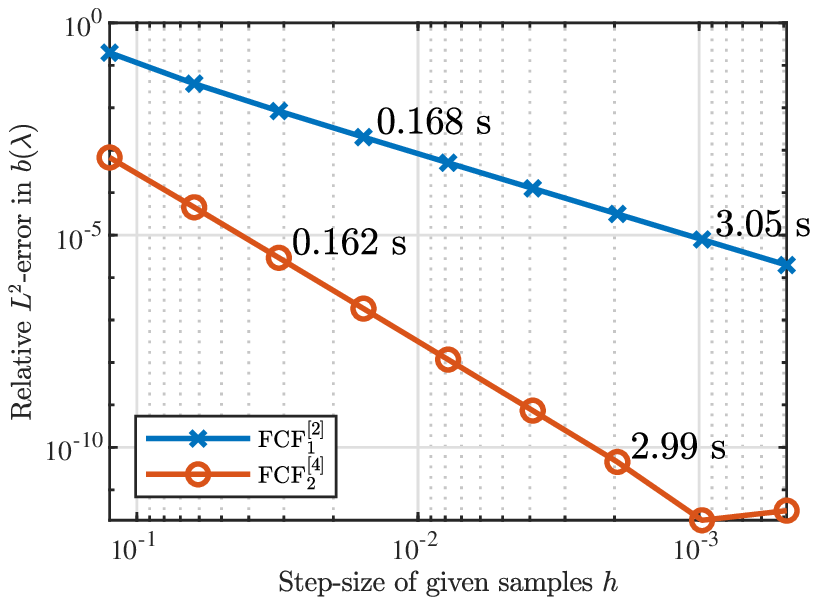}
{Error in $b$-coefficient using FCF algorithms for Example 1. The execution times for some of the points are shown to give an indication of the computational complexity.
\label{fig:FCF_sech_focusing_error_b}}

From the results of the numerical tests presented in this section it is clear that approximating $H(\lambda)$ in \eqref{eqn:Tmatrix} using rational functions to make the method fast, provides a significant computational advantage: similar accuracy, shorter execution time. However, as mentioned earlier we could only make the fourth-order method $\text{CF}_2^{[4]}$ fast. To further improve the accuracy and order of convergence while restricting ourselves to a fourth-order method, we explore the possibility of using Richardson extrapolation in the next section.

\section{Main Result: Fast Sixth-order NFT}\label{Series_acceleration}
In this section we arrive at our main result by integrating Richardson extrapolation into our new fast fourth-order NFT $\text{FCF}_2^{[4]}$ from the previous section. \textbf{We show numerically that the resulting algorithm has \emph{sixth-order} accuracy rather than fifth-order as would be expected. We furthermore show that the added complexity due to Richardson extrapolation is outweighed by the gain in accuracy so the complexity-accuracy trade-off of our final algorithm is the best among all methods investigated in this paper.}

\subsection{Richardson Extrapolation}\label{RE}
Richardson extrapolation is a technique for improving the rate of convergence of a series \cite{Extrapolation_methods}.\footnote{It was used to improve an inverse NFT algorithm for the defocusing case in \cite{Choi2013}.} Given an $r^{\text{th}}$-order numerical approximation method $\hat{\rho}(\lambda, h)$ for the reflection coefficient $\rho(\lambda)$ that depends on the step-size $h$, we can write 
\begin{equation}
\rho(\lambda) = \hat{\rho}(\lambda,h) + \mathcal{O}(h^{r}).
\end{equation}
We assume that $\hat{\rho}(\lambda,h)$ has a series expansion in $h$,
\begin{equation}
\hat{\rho}(\lambda,h) = \rho(\lambda) + \rho_r(\lambda)h^r + \rho_{r+1}(\lambda)h^{r+1} + \dots
\label{eqn:series_expansion}
\end{equation}
In Richardson extrapolation \cite{Extrapolation_methods}, we combine two numerical approximations $\hat{\rho}(\lambda,h)$ and $\hat{\rho}(\lambda,2h)$ as follows,
\begin{equation}
\hat{\rho}^{[\text{RE}]}(\lambda,h) = \frac{2^{r}\hat{\rho}(\lambda,h)-\hat{\rho}(\lambda,2h)}{2^{r}-1}.
\label{eqn:RE}
\end{equation}
Using the series expansion, we find that the order of the new approximation $\hat{\rho}^{[\text{RE}]}(\lambda,h)$ is at least $r+1$:
\begin{equation}
\begin{aligned}
\hat{\rho}^{[\text{RE}]}(\lambda,h) &= \frac{2^r(\rho(\lambda)+\rho_r(\lambda) h^r+\rho_{r+1}(\lambda) h^{r+1}+\dots)}{2^r-1} \\
& - \frac{\rho(\lambda)+\rho_r(\lambda) (2h)^r+\rho_{r+1}(\lambda) (2h)^{r+1}+\dots}{2^r-1} \\
& = \rho(\lambda) - \frac{2^r}{2^r-1}\rho_{r+1}(\lambda)h^{r+1} + \mathcal{O}(h^{r+2}).
\end{aligned}
\end{equation}

We apply this idea to $\text{FCF}_1^{[2]}$ and $\text{FCF}_2^{[4]}$ to obtain the algorithms $\text{FCF\_RE}_1^{[2]}$ and $\text{FCF\_RE}_2^{[4]}$ respectively. Note that the range of $\lvert \text{Re}(\lambda)\rvert$ that can be resolved is determined by the larger of the two step-sizes $h$ (see Section \ref{fast_scattering}). We also remark that Richardson extrapolation can also be applied to the slow algorithms in Section \ref{sec:CFQM}. 
\subsection{Numerical examples}
We test $\text{FCF\_RE}_1^{[2]}$ and $\text{FCF\_RE}_2^{[4]}$ against $\text{FCF}_1^{[2]}$ and $\text{FCF}_2^{[4]}$ for all three examples. Since Richardson extrapolation requires us to compute two approximations, which increases the computational complexity, we again evaluate the complexity-accuracy trade-off. We plot the error versus execution time curves for the three examples in the Figs. \ref{fig:FCFvsFCF_RE_sech_focusing_tradeoff} to \ref{fig:FCFvsFCF_RE_sech_defocusing_tradeoff}. In all figures we can see that the FCF\_RE algorithms achieve slopes of $r+2$ rather than the expected slope of $r+1$. This is an example of superconvergence \cite{Superconvergence}. Specifically, the error of $\text{FCF\_RE}_1^{[2]}$ decreases with slope four and that of $\text{FCF\_RE}_2^{[4]}$ decreases with slope six. As seen before in Section \ref{sec:Numerical_examples}, the arithmetic error starts to dominate after a certain point and causes the error to rise. Although the executions times of FCF\_RE algorithms are higher for the same step-size $h$, the error achieved is almost an order of magnitude lower even for large $h$. From the other viewpoint, for the same execution time, the FCF\_RE algorithms achieve significantly lower errors compared to FCF algorithms. $\text{FCF\_RE}_2^{[4]}$ outperforms $\text{FCF\_RE}_1^{[2]}$ in the trade-off for all the three examples again highlighting the advantage of using higher-order CFQM exponential integrators. These results suggest that Richardson extrapolation should be applied to improve the considered FNFT algorithms. The $\text{FCF\_RE}_2^{[4]}$ algorithm provides the best trade-off among all the algorithms presented in this paper.  

\Figure[t][width=3in,height=2.5in]{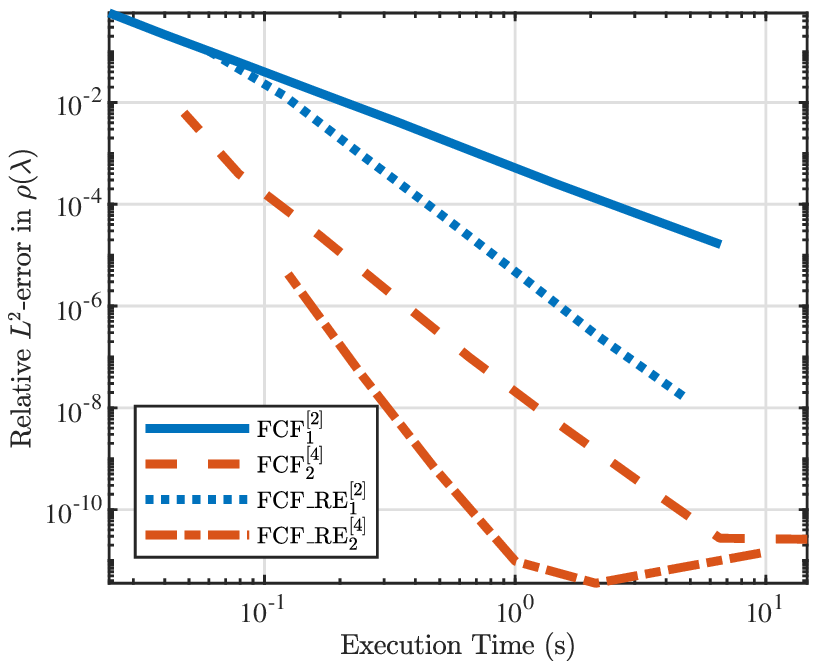}
{Error vs. Execution time trade-off of FCF and FCF\_RE algorithms for Example 1.
\label{fig:FCFvsFCF_RE_sech_focusing_tradeoff}}

\Figure[t][width=3in,height=2.5in]{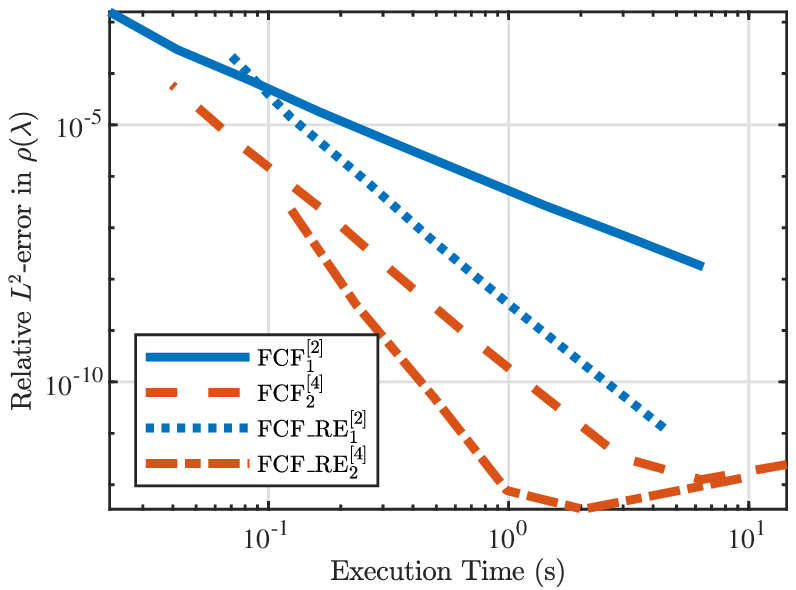}
{Error vs. Execution time trade-off of FCF and FCF\_RE algorithms for Example 2.
\label{fig:FCFvsFCF_RE_rational_1_pole_tradeoff}}

\Figure[t][width=3in,height=2.5in]{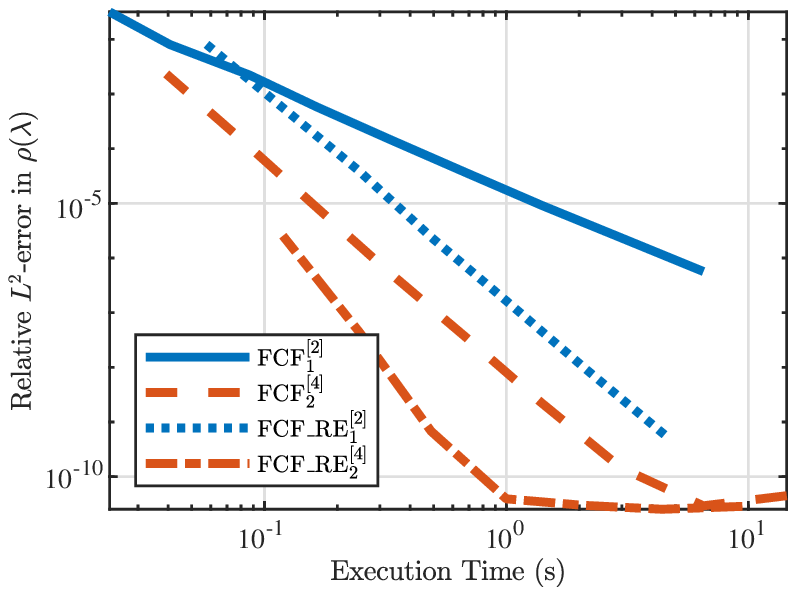}
{Error vs. Execution time trade-off of FCF and FCF\_RE algorithms for Example 3.
\label{fig:FCFvsFCF_RE_sech_defocusing_tradeoff}}

\subsection{Remarks on computing eigenvalues}
The main focus of this paper has been the efficient computation of the reflection coefficient. The computation of the discrete spectrum (see Section \ref{Preliminaries}) is more involved and problem specific. The best approach strongly depends on the available a priori knowledge on the number and location of the eigenvalues. In scenarios where little a priori knowledge is available, some of the ideas presented for the reflection coefficient can be applied to the discrete spectrum as well. Some possible approaches are discussed in Appendix C.

\section{Conclusion}\label{conclusion}
In this paper, we proposed new higher-order nonlinear Fourier transform algorithms based on a special class of exponential integrators. We also showed that one of these algorithms can be made fast using special higher-order exponential splittings. The accuracy of the fast algorithm was improved even further, to sixth-order, using Richardson extrapolation.\textbf{ To the best of our knowledge this is the first fast sixth-order NFT algorithm ever presented in the literature.} Numerical demonstrations showed that the proposed algorithm is highly accurate and provides much better complexity-accuracy trade-offs than existing algorithms. In the future we plan to integrate the algorithms from this paper into the open source software library FNFT \cite{software}. We finally remark that the development of a fast higher-order inverse NFT is an interesting open topic for future research. \appendices

\appendices 

\section{Comparison of FLOP counts and execution times}\label{appendix_1}
In this section we show a comparison between the number of floating-point operations (FLOPs) and execution times of two algorithms proposed in this paper. We counted all the operations of the slow algorithm and the fast algorithm based on the CF$^{[4]}_2$ integrator. We list the different operations and how often they occur in Table \ref{tab1}. For the FFT and CZT, the size of the input is also specified. The number of FLOPs required for each operation type are provided in Table \ref{tab2}. Note that these values are only rules of thumb and vary widely across programming languages and CPU architectures. The number of FLOPs required for the fast scattering step (see Section \ref{fast_scattering}) is given by
\begin{equation}
\begin{aligned}
&\FLOPs(\text{Fast scattering of size N})\\
&=\sum_{k=0}^{\ceil{\log_2N}} 2^{\ceil{\log_2N}-k}\Big(12\FLOPs(\text{FFT of size}(2^{k+1}+1))\\&+
(8\FLOPs(\text{Mult})+4\FLOPs(\text{Add}))(2^{k+1}+1)\Big).
\end{aligned}
\label{eqn:fpoly}
\end{equation}
In Fig. \ref{fig:FLOPs_sech_focusing}, we plot the total number of FLOPs and the execution times from our MATLAB implementations against the number of given samples $D$. At medium to high number of samples we see that the MATLAB execution times match the number of FLOPs very well. Moreover the crossover point at which the fast algorithm becomes faster than the slow variant ($D>300$ in Fig. \ref{fig:FLOPs_sech_focusing}) is almost the same. At lower number of samples, the execution times deviate from the number of FLOPs. This is due to the unaccounted overheads dominating over the floating-point operations. 
\begin{table}[h]\centering
	\caption{Number of operations per type}
	\setlength{\tabcolsep}{3pt}
	\begin{tabular}{|M{70pt}|M{70pt}|M{70pt}|} 
		\hline
		\multirow{2}{*}{Operation}&\multicolumn{2}{M{140pt}|}{Algorithm}\\\cline{2-3}
		&
		CF$^{[4]}_2$& 
		FCF$^{[4]}_2$ \\
	    \hline
		FFT (no, sz)& $3$, $D$ &$3$, $D$\\
		Multiplication& $14D+9M+18DM$&$86D+4(M+1)$\\
		Addition&$4(D+1)+10DM$ &$34D+6$\\
		Division&$2DM$ &$24D+M$\\
		Conjugation & $2D$ &$2D$\\
		Square-root& $2DM$ &$2D$\\
		sinh & $2DM$ & -\\
		cosh & $2DM$ & -\\
		cos &- & $4D$\\
		sinc &- & $4D$\\
		Exponential & $2M$ &$2M$\\
		CZT (no, sz) & - &$2$, $4D+1$\\
		Fast scattering (no,sz) & - & $1$, $2D$\\
		\hline
		\multicolumn{3}{m{225pt}}{The abbreviations no and sz are short for number and size respectively. All operations are assumed to be on complex numbers. The number of signal samples ($D$) is assumed to be greater than the number of reflection coefficient samples ($M$) being computed, i.e. $D\geq M$.}
	\end{tabular}
	\label{tab1}
\end{table}

\begin{table}[h]\centering
	\caption{Number of FLOPs for various operations}
	\label{tab2}
	\setlength{\tabcolsep}{3pt}
	\begin{tabular}{|M{100pt}|M{100pt}|} 
		\hline
		Operation& 
		Number of FLOPs \\
		\hline
		FFT of size $N$& $5N\log_2N$\\
		Multiplication& $1$\\
		Addition&$1$\\
		Division&$4$\\
		Conjugation & $1$\\
		Square-root& $4$\\
		sinh or cosh& $8$\\
		sin or cos & $8$\\
		sinc &$12$\\
		Exponential & $8$\\
		CZT of size $N$ & $3(5(2N-1)\log_2(2N-1))$\\
		Fast scattering of size N & See \eqref{eqn:fpoly}\\
		\hline
		\multicolumn{2}{m{200pt}}{The number of FLOPs for the basic operations have been taken from \cite[p. 5]{FLOPS_weights}. The number of FLOPs for a sinc are the sum of number of FLOPs for a sin and a division. The number of FLOPs for a FFT are based on the asymptotic number of operations for the radix-2 Cooley-Tukey algorithm	\cite[p. 3]{FFTW3}. The number of FLOPs for a CZT are approximated using three FFTs of size $2N-1$.}
	\end{tabular}
\end{table}

\Figure[t][width=3in,height=2.5in]{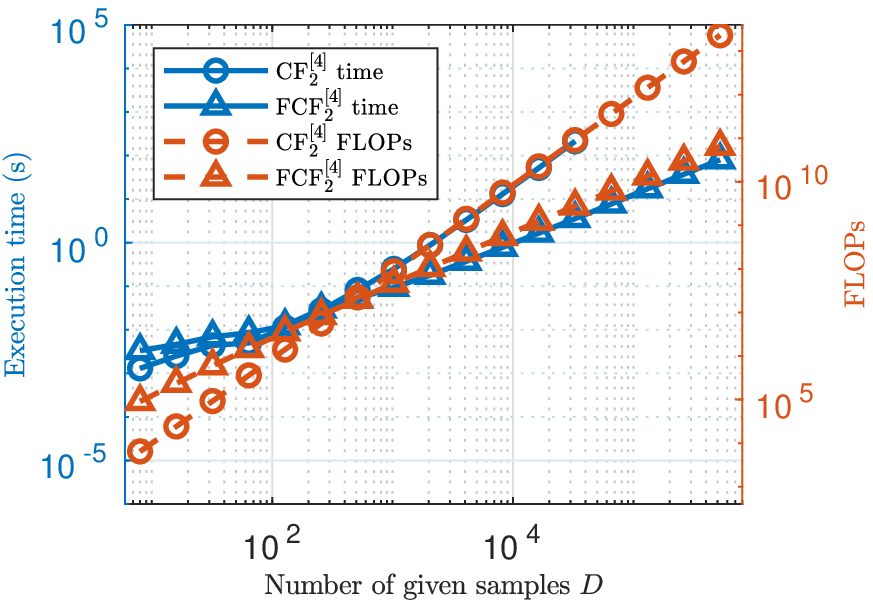}
{The number of FLOPs and measured execution times of a slow fourth-order algorithm and its fast variant.
\label{fig:FLOPs_sech_focusing}}

\section{Interpolation based on Fourier transform}\label{appendix_2}
\begin{lstlisting}[style=Matlab-editor]
function qs = bandlimited_interp(tn,qn,ts)
% Inputs
% tn - Sorted vector of equispaced points at  
%      which qn samples are known
% qn - Vector of signal samples at 
%      points tn
% ts - Value by which samples are
%      to be shifted
% Output 
% qs - Vector of interpolated signal 
%      samples at tn-ts 
ep = tn(2)-tn(1);
Qn = fft(qn);
N = length(qn);
Np = floor(N/2);
Nn = -floor((N-1)/2);
Qn = Qn.*exp(2i*pi*[0:Np,Nn:-1]*ts/(N*ep));
qs = ifft(Qn);
end
\end{lstlisting}

\section{Computing Eigenvalues}\label{DS}

Recall that for the case of focusing NSE ($\kappa=1$), the nonlinear Fourier spectrum has two parts: a continuous and a discrete part. In this appendix, we are concerned with the numerical computation of the discrete part. We first mention some of the existing approaches and then show how one of them can be extended to work with the new fast higher-order NFT algorithms. We will also show that Richardson extrapolation can be applied to improve the accuracy at virtually no extra computational cost.

\subsection{Existing approaches}
Finding the eigenvalues consists of finding the complex upper half-plane roots of the function $a(\lambda)$. Most of the existing approaches can be classified into four main categories.
\begin{enumerate}
	\item \emph{Search methods:} Newton's method.
	\item \emph{Eigenmethods:} Spectral methods based on the solution of a suitably designed eigenproblem \cite{IT_NFT}.
	\item \emph{Gridding methods:}	They find $\lambda_k$ using iterative methods or optimized grid search \cite{IT_NFT,RK4}. Recently a method based on contour integrals was proposed \cite{contour_integral}. 
	\item \emph{Hybrid methods:} Any combination of the above. Eigenmethods with rougher sampling can e.g. be used to find initial guesses for a search method \cite{Subsample_refine}.
\end{enumerate}
Our proposed method will be a hybrid of a eigenmethod and a search method in the spirit of \cite{Subsample_refine}, where an eigenproblem is solved to obtain initial guesses for Newton's method.
\subsection{Proposed method}
Remember that the discrete spectrum consists of eigenvalues, which are the zeros of ${a}(\lambda)$ in the complex upper half-plane ($\mathbb{H}$), and their associated residues. We start with discussing an approximation of ${a}(\lambda)$ that will be useful for locating the eigenvalues.
From \eqref{eqn:Jost}, \eqref{eqn:Scattering_matrix} and \eqref{eqn:NFT_AKNS} we can write
\begin{equation}
a(\lambda) = \lim\limits_{t\to \infty}\phi_1(t,\lambda)e^{i\lambda t}.
\end{equation}
Over the finite interval $[T_-,T_+]$ using \eqref{eqn:Tmatrix} we can see that 
\begin{equation}
a(\lambda) \approx H_{1,1}(\lambda)e^{i\lambda T_-}e^{i\lambda T_+}.
\end{equation}
Hence we hope that the zeros of $H_{1,1}(\lambda)$ are approximations of the zeros of $a(\lambda)$ if the signal truncation and discretization errors are small enough. In Section \ref{fast_scattering} we explained how $H(\lambda)$ can be approximated by a rational function in a transformed coordinate $z$. Hence we can further write
\begin{equation}
a(\lambda) \approx \frac{\hat{a}_{\text{num}}(z)}{\hat{a}_{\text{den}}(z)}e^{i\lambda T_-}e^{i\lambda T_+},
\end{equation} 
where $\hat{a}_{\text{num}}(z)$ and $\hat{a}_{\text{den}}(z)$ are polynomials in $z(\lambda)$. Let $\hat{a}_{\text{num}}(z)=\hat{a}_Nz^N+\hat{a}_{N-1}z^{N-1}+\ldots+\hat{a}_0$. Thus $\hat{a}_{\text{num}}(z)$ will have $N$ zeros. These zeros or roots of $\hat{a}_{\text{num}}(z)$ can be found using various methods \cite{root_finding}. Of these $N$ zeros, there should be $K$ (typically, $K\ll N$) values that are approximations of zeros of $a(\lambda)$ in $\mathbb{H}$. 
\subsubsection*{Example}
We would like to add clarity through a visual representation of the roots. We choose the signal from Example 1 with $D = 2^9$. We plot all the zeros of $\hat{a}_{\text{num}}(z)$ of $\text{FCF}_1^{[2]}$ with `x' in Fig. \ref{fig:FCF2_1_a_z}. Here $z = e^{i\lambda h}$. We can then map these zeros back to obtain values of $\lambda$. These are plotted with `x' in Fig. \ref{fig:FCF2_1_a_lam}. From the definition of discrete spectrum, we can filter out all the values that are not in $\mathbb{H}$. Recall that we can resolve the range $\lvert \text{Re}(\lambda)\rvert<\pi/h$. (See Section \ref{fast_scattering}.) Since we observed that spurious eigenvalues tend to cluster around $\lvert \text{Re}(\lambda)\rvert=\pi/h$, we filter out the corresponding roots of $\hat{a}_{\text{num}}(z)$. More precisely we keep only values of $\lambda$ for which $\lvert \text{Re}(\lambda)\rvert<0.9\pi/h$. The filtered roots are plotted in Figs. \ref{fig:FCF2_1_a_z} and \ref{fig:FCF2_1_a_lam} with `o'. For the chosen value of $\mathring{q}=5.4$ the set of eigenvalues is $\Lambda=\{3+4.9i,3+3.9i,3+2.9i,3+1.9i,3+0.9i\}$. From Fig. \ref{fig:FCF2_1_a_lam} we see that the values marked with `o' are indeed approximations of the values in set $\Lambda$. However, there is no guarantee that we will always be able to locate approximations for all values in $\Lambda$ as that depends on several factors, some of which are the signal magnitude $q_o$, signal interval $[T_-,T_+]$, step-size $h$ and values of the eigenvalues themselves. 

For the example chosen in the visual demonstration, the \text{num}ber of zeros is $N=1024$ and the number of eigenvalues is $K=5$. For the chosen mapping from $\lambda \to z$, the $K$ values of interest will always lie inside the unit circle in Fig. \ref{fig:FCF2_1_a_z} and most other spurious zeros of $\hat{a}_{\text{num}}(z)$ will lie on the unit circle. Even with the best eigenmethods available for polynomial root-finding, which have a complexity of $\mathcal{O}(N^2)$ \cite{fast_root_finder}, execution time grows very steeply, making this approach infeasible for large $N$. To reduce the complexity, it was suggested in \cite{Subsample_refine} to sub-sample the given signal to reduce the dimensionality of the root-finding problem. The algorithm is summarized in Fig. \ref{alg1}.

\Figure[t][width=3in,height=2.5in]{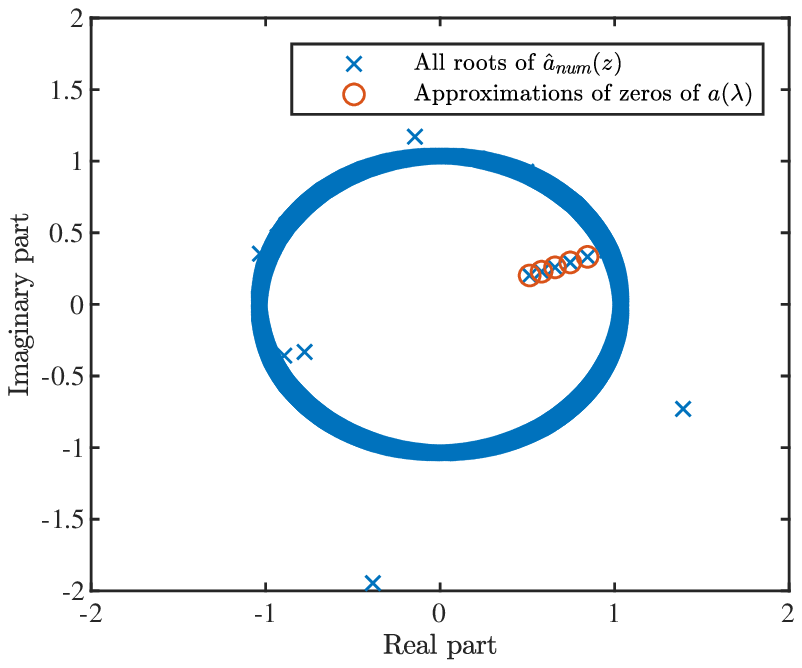}
{Zeros of $\hat{a}_{\text{num}}(z)$ for Example 1 with $D=2^9$.
	\label{fig:FCF2_1_a_z}}

\Figure[t][width=3in,height=2.5in]{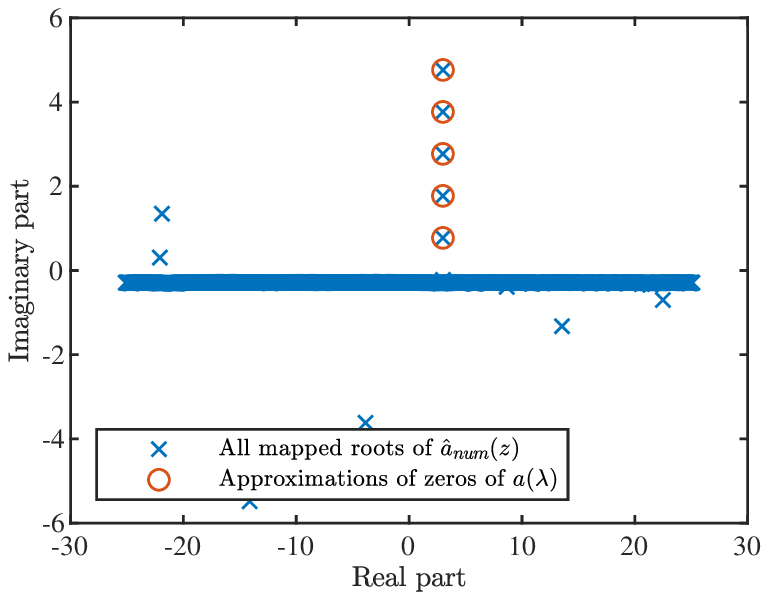}
{Mapped zeros of $\hat{a}_{\text{num}}(z)$ for Example 1 with $D=2^9$.
	\label{fig:FCF2_1_a_lam}}

\begin{figure}[t]
	
	\begin{algorithmic}[1]
		\renewcommand{\algorithmicrequire}{\textbf{Input:}}
		\renewcommand{\algorithmicensure}{\textbf{Output:}}
		\REQUIRE Samples $q_0,\dots,q_{D-1}$, $D_{\text{sub}}$, $[T_-,T_+]$
		\ENSURE  Estimated eigenvalues $\hat{\lambda}_k$
		\STATE Subsample the given samples $\rightarrow q_{n}^{\text{sub}}:=q_{n\lfloor D/D_{\text{sub}} \rceil}$, where $\round{\cdot}$ rounds to the closest integer.\\
		Build a polynomial approximation $\hat{a}_{\text{num}}(z)$ using the $D_{\text{sub}}$ subsampled samples.
		\\Apply a fast eigenmethod to find the roots of $\hat{a}_{\text{num}}(z)$.
		\\Filter the roots.
		\STATE Refine the roots via Newton's method using all samples. \\
		Filter the refined roots.
		\STATE Apply Richardson extrapolation to the unrefined and refined roots.
	\end{algorithmic}
	\caption{\textbf{Algorithm :} Subsample and refine \label{alg1}}
\end{figure}

We now discuss the three stages of the algorithm in detail.
\begin{enumerate}
	\item \textbf{Root finding from a subsampled signal}\\
	The given signal $q_n$ is subsampled to give $q_{n}^{\text{sub}}$ with $D_{\text{sub}}$ samples. The corresponding step-size is $h_{\text{sub}}$. There are no results for the minimum number of samples that guarantee that all eigenvalues will be found. One choice can be based on limiting the overall computational complexity to $\mathcal{O}(D\log^2D)$, which is the complexity for the reflection coefficient. For a root-finding algorithm with $\mathcal{O}(D^2)$ complexity, we choose to use $D_{\text{sub}}=\text{round}\big(\sqrt{D\log^2D}\big)$ samples. The polynomial $\hat{a}_{\text{num}}(z)$ is then built from these $D_{\text{sub}}$ samples. For $\text{FCF}_2^{[4]}$, the non-equispaced samples should be obtained from the original $D$ samples and not the $D_{\text{sub}}$ samples. An eigenmethod is then used to find all zeros of $\hat{a}_{\text{num}}(z)$. We used the algorithm in \cite{fast_root_finder}. The values of $z$ are mapped backed to $\lambda$ and filtered to remove implausible values. 
	\item\textbf{Root refinement using Newton method}\\
	The Newton method based on the slow CF methods is used for root refinement. The derivative $\mathrm{d} a(\lambda)/\mathrm{d} \lambda$ is calculated numerically along with $a(\lambda)$ as in \cite{BO} using all the given samples $q_n$. The values of $\lambda$ that remain after filtering in the previous step are used as the initial guesses for the Newton method. We chose to stop the iterations if the change in value goes below $10^{-15}$ or if a maximum of 15 iterations is reached. The resulting roots are filtered again.
	\item \textbf{Richardson extrapolation}\\
	We pair the roots resulting from the Newton step, $\hat{\lambda}_k^{\text{Newton}}$, with the corresponding initial guesses $\hat{\lambda}_k^{\text{init}}$. Then, we apply Richardson extrapolation:
	\begin{equation}
	\hat{\lambda}_k = \frac{(\frac{h_{\text{sub}}}{h})^{r}\hat{\lambda}_k^{\text{Newton}}-\hat{\lambda}_k^{\text{init}}}{(\frac{h_{\text{sub}}}{h})^{r}-1}.
	\end{equation}
	$\hat{\lambda}_k$ is then an improved approximation and constitutes the discrete part of the FCF\_RE algorithm. It may so happen that more than one $\hat{\lambda}_k^{\text{init}}$ converge to the same $\hat{\lambda}_k^{\text{Newton}}$. In such cases the value $\hat{\lambda}_k^{\text{init}}$ closest to $\hat{\lambda}_k^{\text{Newton}}$ should be used for Richardson extrapolation. The other values $\hat{\lambda}_k^{\text{init}}$ that also converged to the same $\hat{\lambda}_k^{\text{Newton}}$ should be treated as spurious values and discarded.
\end{enumerate}

The numerical algorithms may not find particular eigenvalues or find spurious ones. Let $\hat{\Lambda}$ be the set of approximations found by an algorithm. To penalize both missed values and incorrect spurious values at the same time, we define the error
\begin{equation}
\begin{aligned}
E\Lambda=\maxi\Big\{
&\maxi_{\vphantom{ \hat{\lambda}_j\in\hat{\Lambda}}   \lambda_i\in\Lambda} \mini_{ \hat{\lambda}_j\in\hat{\Lambda}} \abs{\lambda_i-\hat{\lambda}_j},\\
&\maxi_{\hat{\lambda}_j\in\hat{\Lambda}} \mini_{\vphantom{ \hat{\lambda}_j\in\hat{\Lambda}} \lambda_i\in\Lambda}  \abs{\lambda_i-\hat{\lambda}_j}\Big\}.  
\end{aligned}
\end{equation}
Note that the first term in the outer maximum grows large if an algorithm fails to approximate a part of the set $\Lambda$ while the second term becomes large if an algorithm finds spurious values that have no correspondence with values in $\Lambda$. $E\Lambda$ is expected to be of order $r$ for an algorithm of order $r$. 
\subsection{Numerical Example}
In this section, we compare different variants of our proposed algorithm using Example 1. We compute the error $E\Lambda$ for the following three types of algorithms:
\begin{enumerate}
	\item Discrete part of FCF algorithms. An eigenmethod is applied to the approximation $\hat{a}_{\text{num}}(z)$ built using all samples. No sub-sampling is used.
	\item Discrete part of FCF algorithms with sub-sampling. Only steps 1 and 2 of the algorithm mentioned above.
	\item Discrete part of FCF\_RE algorithms. All the three steps mentioned above.
\end{enumerate}
To demonstrate the effect of sub-sampling, we show in Fig. \ref{fig:FCF_sech_focusing_subsample_refine_discspec_error} the errors for the second- and fourth-order algorithms of types 1 and 2. For $h>0.3$ the errors are high either due to failure to find approximations close to the actual eigenvalues or due to spurious values. For $h\leq0.3$, $\text{FCF}_2^{[4]}$ of type 1 and $\text{FCF}_1^{[2]}$ of type 2 find exactly five values that are close approximations of the values in $\Lambda$. However $\text{FCF}_1^{[2]}$ of type 1 and $\text{FCF}_2^{[4]}$ of type 2 find good approximations only for $h\leq0.06$. The error of $\text{FCF}_1^{[2]}$ algorithms decreases with slope two and that of $\text{FCF}_2^{[4]}$ algorithms decreases with slope four as expected from the order of the underlying numerical schemes.   

In Fig. \ref{fig:FCF_sech_focusing_subsample_refine_discspec__RE_error} we show the errors for the second- and fourth-order algorithms of type 2 and 3 to indicate the advantage of the extrapolation step. The extrapolation step improves the approximation significantly for $\text{FCF\_RE}_1^{[2]}$ while adding negligible computation cost to the algorithm. However, there is only minor improvement in case of $\text{FCF\_RE}_2^{[4]}$ over $\text{FCF}_2^{[4]}$.  

In Fig. \ref{fig:FCF_sech_focusing_subsample_refine_discspec_time} we plot the execution times for the $\text{FCF}$ algorithms of types 1 and 2. The execution times of algorithms of type 3 are almost the same as those of type 2. For type 1 algorithms, these times include the time required to build $\hat{a}_{\text{num}}(z)$ and the time taken by the root-finder. For algorithms of type 2, the additional time required for root-refinement by Newton's method is also included. Even with sub-sampling, we see that the execution times are an order of magnitude higher than the execution times for the continuous part. The $\text{FCF\_RE}$ algorithms seem to provide the best trade-off between accuracy and computation cost similar to the case of continuous part. The overall computational complexity may be decreased by using alternative methods to find the initial guesses.

\Figure[t][width=3in,height=2.5in]{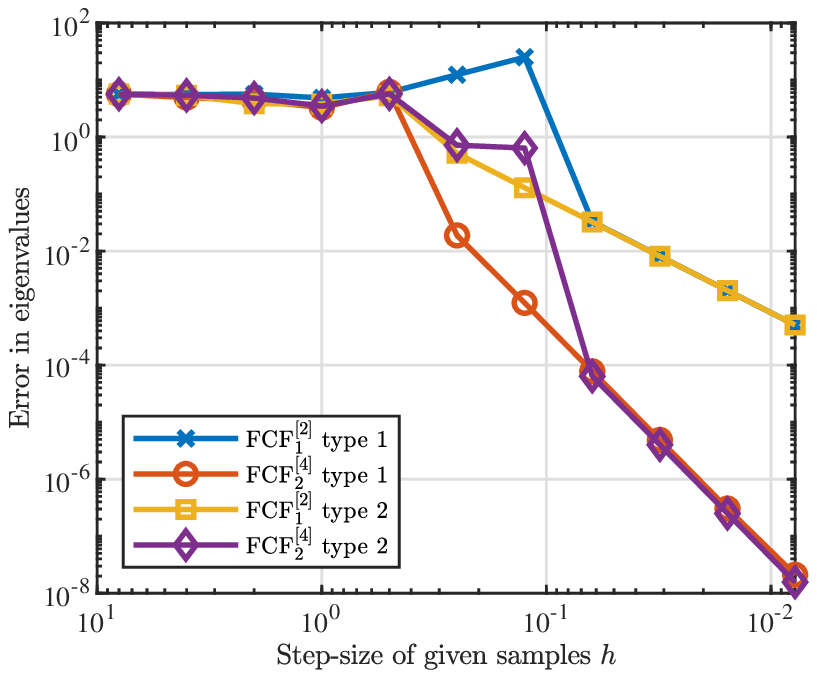}
{Error in approximation of the eigenvalues by the fast second- and fourth-order algorithms of type 1 (no sub-sampling) and type 2 (sub-sample and refine, no Richardson extrapolation).
	\label{fig:FCF_sech_focusing_subsample_refine_discspec_error}}

\Figure[t][width=3in,height=2.5in]{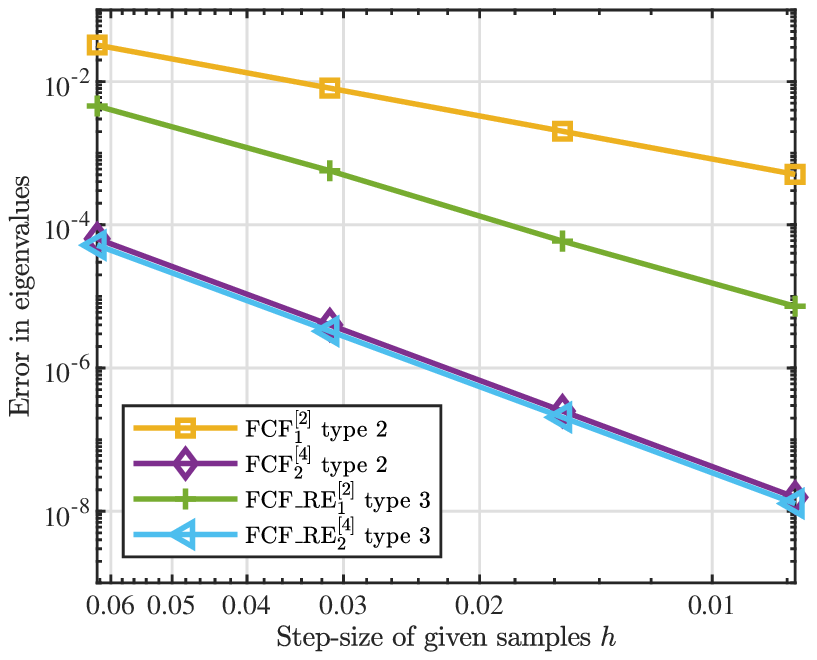}
{Error in approximation of a eigenvalues computed using FCF and FCF\_RE algorithms.
	\label{fig:FCF_sech_focusing_subsample_refine_discspec__RE_error}}

\Figure[t][width=3in,height=2.5in]{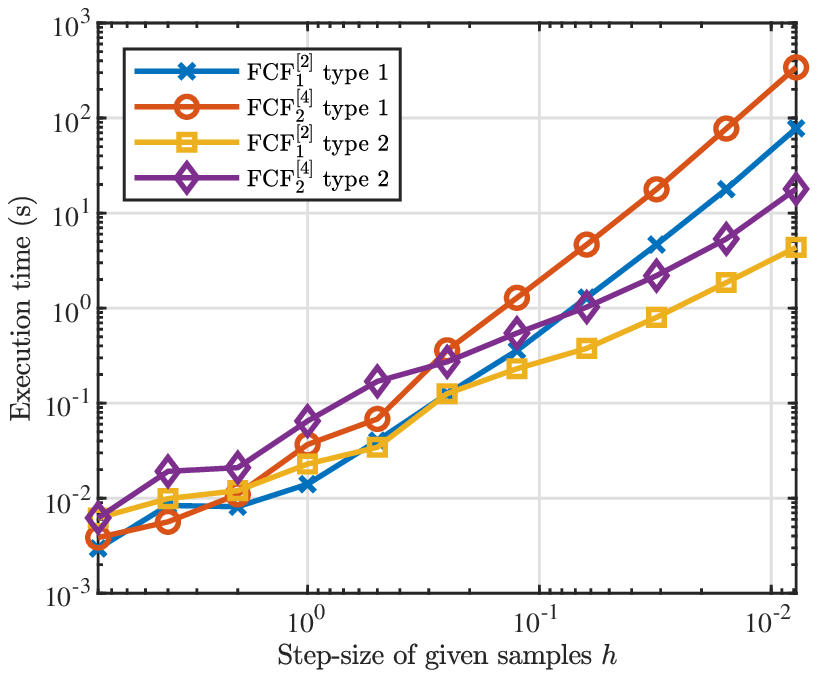}
{Execution time of $\text{FCF}_2^{[4]}$ and $\text{FCF}_2^{[4]}$ algorithms for computing eigenvalues of Example 1.
	\label{fig:FCF_sech_focusing_subsample_refine_discspec_time}}

\bibliography{MyBib}
\bibliographystyle{IEEEtranDOI}

\EOD

\end{document}